\documentclass[twocolumn,trackchanges]{aastex631}

\usepackage{dirtytalk}

\begin{document}


\title{Stellar Rotation and Structure of the $\alpha$ Persei Complex \\
\small When Does Gyrochronology Start to Work?}

\correspondingauthor{Andrew Boyle}
\email{awboyle@caltech.edu}

\author[0000-0001-6037-2971]{Andrew W. Boyle}
\affiliation{Department of Astronomy, California Institute of Technology, 1200 E. California Blvd, Pasadena, CA 91125}

\author[0000-0002-0514-5538]{Luke G. Bouma}
\altaffiliation{51 Pegasi b Fellow}
\affiliation{Department of Astronomy, California Institute of Technology, 1200 E. California Blvd, Pasadena, CA 91125}



\begin{abstract}

On the pre-main-sequence, the rotation rates of Sun-like stars are dictated by the interplay between the protostellar disk and the star's contraction. At ages exceeding 100 million years (Myr), magnetic spin-down erases the initial stellar spin rate and enables rotation-based age dating (gyrochronology). The exact time at which the transition between these two regimes occurs depends on stellar mass, and has been challenging to empirically resolve due to a lack of viable calibration clusters. The $\alpha$ Persei open cluster ($t\approx80$ Myr, $d\approx170$ pc) may provide the needed calibrator, but recent analyses of the Gaia data have provided wildly varying views of its age and spatial extent. As such, we analyze a combination of TESS, Gaia, and LAMOST data to calibrate gyrochronology at the age of $\alpha$ Per and to uncover the cluster's true morphology. By assembling a list of rotationally-confirmed $\alpha$ Per members, we provide strong evidence that $\alpha$ Per is part of a larger complex of similarly-aged stars. Through kinematic back-integration, we show that the most diffuse components of $\alpha$ Per were five times closer together 50 Myr ago. Finally, we use our stellar rotation periods to derive a relative gyrochronology age for $\alpha$ Per of 67 $\pm$ 12\% the age of the Pleiades, which yields 86 $\pm$ 16 Myr given current knowledge.  We show that by this age, stars more massive than $\approx$0.8 M$_{\odot}$ have converged to form a well-defined slow sequence.

\end{abstract}



{\catcode`\&=11
\gdef\meingast{\cite{2021A&A...645A..84M}}}

{\catcode`\&=11
\gdef\cg{\cite{2018A&A...618A..93C}}}

{\catcode`\&=11
\gdef\lodieu{\cite{2019A&A...628A..66L}}}

\section{Introduction} \label{sec:intro}

Star clusters form when molecular clouds undergo gravitational collapse. As the clouds collapse, they fragment into clumps and filaments of gas that eventually form stars \citep[e.g.,][]{2014prpl.conf...27A}. During this fragmentation process, stars often form in groups of tens to thousands of other stars that are embedded within the cloud \citep{2003ARA&A..41...57L}. As time passes, ionizing radiation, stellar winds, radiation pressure, and supernova shocks expel the remaining dust and gas, resulting in a loosely bound open cluster \citep{2019ARA&A..57..227K}. 

Observational and theoretical evidence support a hierarchical view of star formation. In the hierarchical view, the spatial distribution of stars inherits substructure from the parent cloud, with clusters being just one outcome of star formation in the densest regions  \citep{2021MNRAS.506.3239G}. The highly non-uniform spatial distribution of stars in young groups such as Sco-Cen \citep{2018MNRAS.476..381W}, the Orion complex \citep{2018AJ....156...84K}, and the Taurus complex \citep{2017ApJ...838..150K, 2021AJ....162..110K} are the nearest clear consequences of hierarchical star formation. More distant examples, such as the $h$ and $\chi$ Persei double cluster \citep{2021ApJ...909...90D} and the W3/4/5 regions \citep{2000ApJS..130..381C}, show that hierarchical star formation is widespread throughout the Galaxy. 

After gas dispersal, the initial configuration of a cluster evolves due to a combination of stellar and galactic dynamics.  The outcome at any given time is therefore a product of the initial stellar locations, kinematics, and any subsequent dynamical processing.   A few dispersing groups that show the importance of both initial conditions and subsequent dynamics include the Sco-Cen complex \citep{2021ApJ...917...23K}, the diffuse populations around the Pleiades, IC 2602, Platais 8, and Octans \citep{2021ApJ...915L..29G}, and the Cep-Her complex \citep{2022AJ....163..121B}.

Disentangling which observed substructures are primordial and which are a consequence of dynamical evolution is challenging because many processes contribute to a dissolving cluster's structure.  For example, two-body relaxation, combined with the differential rotation of the Galaxy, drives the formation of leading and trailing tidal tails \citep[e.g.,][]{2008gady.book.....B,2020A&A...640A..84D, 2020A&A...640A..85D}.  However a separate process that can form diffuse populations around a cluster's core is the star formation process itself, especially if the parent cloud was already filamentary \citep{2018ApJ...864..153Z}.  As an additional complicating factor, a single high-speed encounter between an open cluster and a molecular cloud can remove a significant fraction of the cluster's binding energy \citep{1958ApJ...127...17S,ryden2016dynamics}. The structure of a dissolving cluster is therefore dictated by a combination of these effects \citep[e.g.,][]{2010MNRAS.409..305L}, and unambiguous kinematic signatures of each process are needed in order to untangle them.

A separate challenge is identifying stars in diffuse populations in the first place.  Tidal tails can extend from hundreds \citep{2021A&A...647A.137J} to even thousands \citep{2022MNRAS.514.3579B} of parsecs from the core of their associated cluster, and so it can be hard to differentiate bona fide tidal tail members from unrelated field stars. Efforts to address this challenge have been aided by the Gaia mission, whose goal is to precisely track the position and motion of $\sim$1 billion stars in the Milky Way. It is only with the arrival of Gaia's unprecedented astrometric precision and completeness that it has become possible to consistently discover diffuse populations in the peripheries of nearby clusters \citep[e.g.,][]{2021A&A...645A..84M,2022MNRAS.tmp.2692B}. Because identifying stellar clusters in Gaia data is a relatively new exercise, many methods have been proposed for determining cluster membership. Techniques include using clustering algorithms such as DBSCAN \citep{10.5555/3001460.3001507}, HDBSCAN \citep{McInnes2017}, and UPMASK \citep{2014A&A...561A..57K}, using Gaussian Mixture Models \citep{2021ApJ...923..129J}, or performing kinematic analyses on a selected population of stars \citep{2021arXiv211004296H}. It is currently unclear which of these methods is the most effective for identifying cluster members and diffuse stellar structures.

One way of determining which clustering method is most effective is by analyzing the rotation periods of candidate cluster members. As stars age, their rotation periods tend to increase as they lose angular momentum due to their magnetized winds \citep{1967ApJ...148..217W, 1972ApJ...171..565S, 2007ApJ...669.1167B, 2008ApJ...687.1264M}. The implication is that stellar rotation can be used to age date stars --- a method known as gyrochronology. As stars contract on the pre-main sequence (PMS), their rotation periods decrease due to conservation of angular momentum. If a disk is present, locking between the star's magnetosphere and the inner disk can inhibit spin-down until the disk disperses, at which time the star will resume spinning down \citep{1991ApJ...370L..39K,2005ApJ...634.1214L}. A solar-mass, solar-metallicity star will take $\approx$40 Myr to arrive on the zero-age main sequence \citep{2016ApJ...823..102C}, after which wind braking becomes the dominant mode of angular momentum loss.

Since all stars in an open cluster are born at roughly the same time, provided their rotation rates are dominated by magnetic braking, their rotation periods should trace a smooth trend in effective temperature at any given age. If a star with an anomalous rotation period is found in an open cluster, we can conclude that said star is likely an interloper and not a true cluster member. 

Open clusters provide the empirical foundation for gyrochronology. Studies of stellar rotation have been performed for $\rho$ Ophiuchus \citep[$\sim$1 Myr;][]{2018AJ....155..196R}, Upper Scorpius \citep[$\sim$8 Myr;][]{2018AJ....155..196R}, Tucana-Horologium \citep[$\sim$40 Myr;][]{2022arXiv221105258P}, $\mu$ Tau \citep[$\sim$62 Myr][]{2020ApJ...903...96G} the Pleiades \citep[$\sim$120 Myr;][]{2016AJ....152..113R},  Pisces-Eridanus \citep[$\sim$120 Myr;][]{2019AJ....158...77C}, Blanco 1 \citep[$\sim$146 Myr;][]{2020MNRAS.492.1008G}, NGC 2516 \citep[$\sim$150 Myr;][]{2021AJ....162..197B}, Praesepe \citep[$\sim$670 Myr;][]{2017ApJ...842...83D, 2021ApJ...921..167R}, the Hyades \citep[$\sim$727 Myr;][]{2019ApJ...879..100D}, NGC 6811 \citep[$\sim$1 Gyr;][]{2019ApJ...879...49C}, NGC 752 \citep[$\sim$1.4 Gyr;][]{2018ApJ...862...33A} and Ruprecht 147 \citep[$\sim$2.7 Gyr;][]{2020ApJ...904..140C}.  These studies have established a universal slow sequence for G-dwarf rotation at $\sim$120 Myr \citep{2020A&A...641A..51F}.   However, no rich calibration clusters at an intermediate age of 70--100 Myr have yet been studied. It is therefore unclear what stellar mass ranges, if any, have converged to form a slow rotation sequence at this earlier time.

Given the current landscape, $\alpha$ Per is a prime cluster to use to calibrate gyrochronology because it is at an age ($\sim$70--100 Myr) where stellar rotation has yet to be explored, while its close proximity ($\sim$170 pc) means that TESS photometry can be exploited to obtain rotation periods for a large numbers of its candidate members.
The guiding questions for this work are as follows:

\begin{enumerate}
    \item How do stars rotate as a function of effective temperature at the age of $\alpha$ Per?
    \item What is the best clustering method to use for identifying diffuse stellar structures based on positions and velocities?
    \item What is the true morphology of $\alpha$ Per?
\end{enumerate}

We first provide an overview of the $\alpha$ Persei cluster (Section~\ref{sec:aper}), and proceed by describing the eight different clustering studies included in our analysis (Section~\ref{sec:sample}). We detail how we derived rotation periods from the TESS data (Section~\ref{sec: methods}), and describe our reddening correction, effective temperature scale, and empirical isochrone age calculations (Section~\ref{sec:reddening}).  These considerations inform our main gyrochronology analysis (Section~\ref{sec:gyro}), which we then use to evaluate false-positive rates from different clustering methodologies (Section~\ref{sec:true positive}), as well as $\alpha$ Per's morphology (Section~\ref{sec:morphology}).   Our closing discussion (Section~\ref{sec:discussion}) touches on topics of interest including white dwarfs in $\alpha$ Per, the cluster's metallicity, interpretation of the cluster's morphology in the broader context of cluster dispersal, and the implications of $\alpha$ Per's rotation sequence for the effectiveness of gyrochronology at young ages.  We conclude in Section~\ref{sec: conclusion}.

\section{$\alpha$ Per} \label{sec:aper}

\begin{figure}[t]
    \centering
    \includegraphics[width=0.45\textwidth]{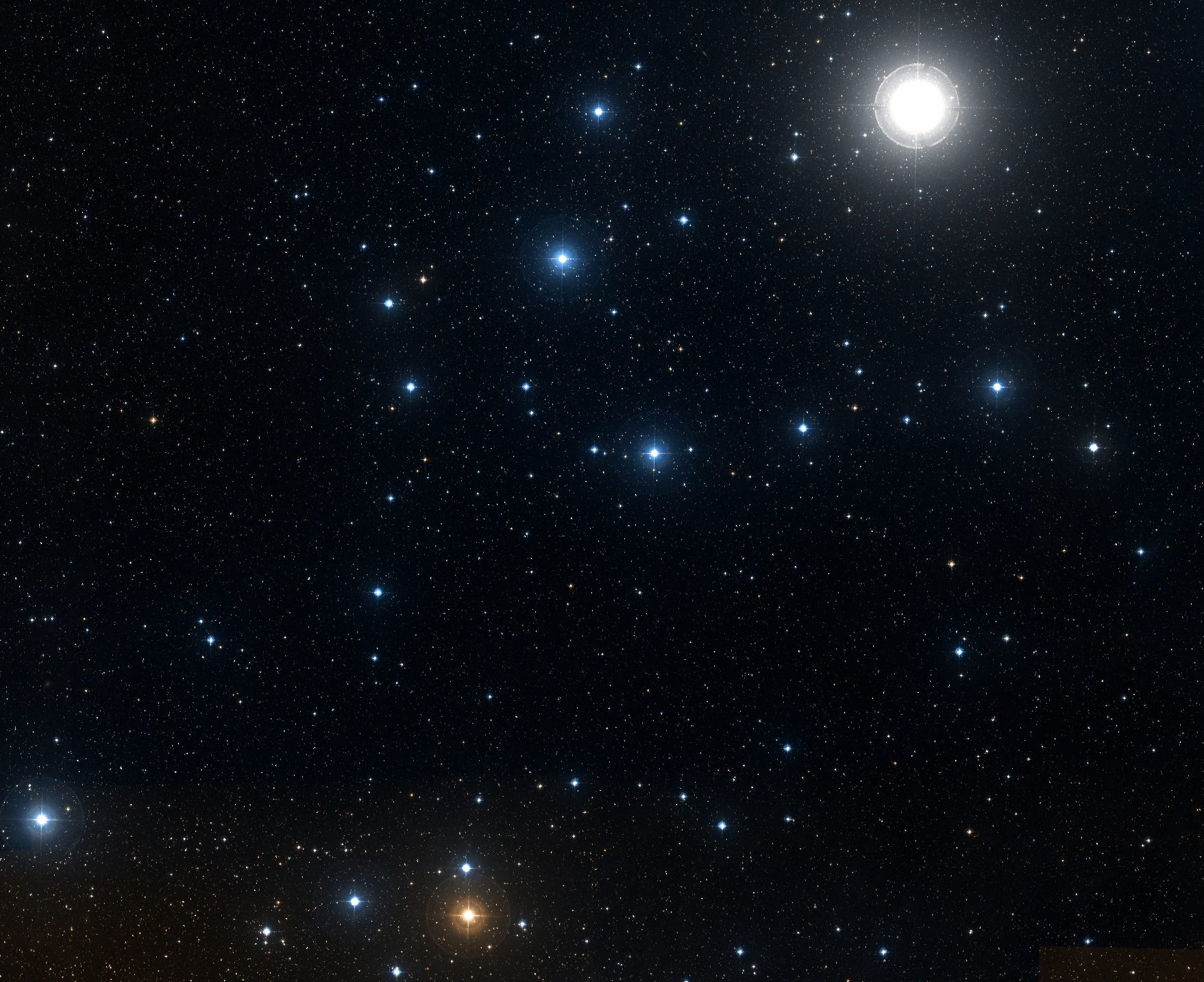}
    \caption{An image of $\alpha$ Per taken from DSS. The image is $\sim$4$^\circ$ across, which corresponds to $\sim$24 parsecs at the distance of the cluster's core. The image is oriented so equatorial North is up and East is left. Reported $\alpha$ Per members in the literature span $\sim$$100^{\circ}$ on the sky, far beyond the extent of this image.}
    \label{fig:aper image}
\end{figure}

The first known description of $\alpha$ Per (Melotte 20, Collinder 39, Theia 133, Crius 229, $l = 147^{\circ}$, $b = -6^{\circ}$, ${\rm age}\sim71$ Myr, $d\sim175$ pc; \citealt{2018A&A...616A..10G}) was reported by Giovanni Batista \citet{1654dsoc.book.....H}, an astronomer at the Duke Court of Montechiaro \citep[see also][]{1985JHA....16....1F}. Hodierna cataloged $\alpha$ Per as a ``\textit{Luminosae}'', or a region containing stars resolvable by the naked eye, likely visible because of its large number of hot, massive stars. $\alpha$ Per is noticeably absent from the Messier Catalog, but was cataloged as a moving cluster by \citet{1910MNRAS..71...43E} and as a large, extended cluster by \citet{1915MmRAS..60..175M}. Figure~\ref{fig:aper image} shows an image of $\alpha$ Per four degrees across and centered on $\alpha$ Per's right ascension and declination. The bright, blue stars that early astronomers used to detect $\alpha$ Per are shown here and demonstrate why $\alpha$ Per is sometimes referred to as an OB association. 

Literature ages for $\alpha$ Per vary, but it is younger than the Pleiades on most age scales. Some representative reported ages include: 51.3 Myr \citep[isochrone analysis;][]{1981A&A....97..235M}, $79.0^{+1.5}_{-2.3}$ Myr \citep[lithium depletion boundary (LDB) measurements;][]{2022A&A...664A..70G}, $90 \pm 10$ Myr \citep[LDB measurements;][]{1999ApJ...527..219S}, and 96--100 Myr \citep[$\delta$ Scuti pulsations;][]{2022MNRAS.513..374P}. The metallicity of $\alpha$ Per similarly has a wide range of reported values (all in terms of [Fe/H]): $-0.054 \pm 0.046$ \citep{1990ApJ...351..467B}, -0.05 \citep{2003AJ....125.1397C}, $0.14 \pm 11$ \citep{2016A&A...585A.150N}, and 0.18 \citep{2010A&A...514A..81P}.

Rotation periods for $\alpha$ Per members have previously been determined using ground-based photometry \citep[e.g.,][]{1987PASP...99..471S, 1989ApJ...346..160S, 1993MNRAS.262..521O, 1993PASP..105.1407P, 1995PASP..107..211P, 1997cfa..rept.....P, 2003ApJ...586..464B, 2008ApJ...687.1264M, 2010A&A...520A..15M, 2015A&A...577A..98G}, but no comprehensive study of $\alpha$ Per's rotation periods has yet been undertaken in the era of space-based light curves. This is partially due to the limitations of ground-based photometry and a lack of large membership lists, but also because $\alpha$ Per's low galactic latitude puts the cluster in a crowded region of the galaxy, making it harder to distinguish bona fide $\alpha$ Per members from unrelated field stars. The combination of TESS's all-sky survey and updated lists of cluster members from Gaia means that $\alpha$ Per's rotation periods can now be explored in a more complete way than was possible in the past.

In summary, $\alpha$ Per's close proximity ($\sim$170 pc), large spatial extent, young age, and substantial population \citep[$\sim$$1000M_{\odot}$,][]{2021A&A...645A..84M} make it an ideal cluster on which to perform a rotation analysis.

\section{Sample Selection} \label{sec:sample}

We select our sample of candidate $\alpha$ Per members based on previously published membership lists. We focus our search on lists derived from studies that use Gaia data due to Gaia’s unprecedented sensitivity, completeness, and precision. After a literature review, we identify eight studies as presenting the most significant membership lists available for $\alpha$ Per: \cite{2018A&A...618A..93C}, \cite{2019AJ....158..122K}, \cite{2019A&A...628A..66L}, \cite{2021arXiv211004296H}, \cite{2021ApJ...923..129J}, \cite{2021ApJ...917...23K}, \cite{2021A&A...645A..84M}, and \cite{2022arXiv220604567M}. Each paper and its method of determining cluster membership is discussed in Appendix \ref{sec: clustering descriptions}.

These studies can be divided into three categories: 

\begin{enumerate}
    \item Studies that used an unsupervised clustering algorithm to perform a blind search on Gaia data \citep{2019AJ....158..122K, 2021ApJ...917...23K, 2022arXiv220604567M}.
    \item Studies that used a clustering algorithm and prior information about the cluster, such as membership lists or previously reported positions and velocities \citep{2018A&A...618A..93C, 2021ApJ...923..129J, 2021A&A...645A..84M}.
    \item Studies that did not use a clustering algorithm but that did incorporate prior information about the cluster into their analysis \citep{2019A&A...628A..66L, 2021arXiv211004296H}.
\end{enumerate}

The effectiveness of each clustering analysis will be analyzed in Section \ref{sec:true positive} and discussed in Section \ref{subsec:cluster discussion}. We also note that although we do not extend our search for $\alpha$ Per candidates beyond the eight papers in this section, \citet{2019A&A...628A..66L} did concatenate $\alpha$ Per membership lists published as far back as 1956. As such, we consider our sample to be relatively complete. The number of candidates from each study is shown in Table \ref{table:counts}.

\begin{deluxetable*}{cCCCCCCC}
\tablecaption{Counts for the number of candidates in our analysis}

\tablehead{
\colhead{Paper} & \colhead{N} & \colhead{N$_{\mathrm{unique}}$} & \colhead{N$_{\mathrm{LC}}$} & \colhead{N$_{\mathrm{Per}}$} & \colhead{N$_{\mathrm{final}}$} & \colhead{True Positive Rate} & \colhead{$G_{\rm limit}$}
}

\startdata
\cg & 873 & 17 & 856 & 806 & 227 & 92.6 \pm 7.2$\%$ & 18\\
\citet{2019AJ....158..122K} & 2643 & 1187 & 2444 & 2286 & 674 & 76.7 \pm 7.7$\%$ & 19\\
\lodieu & 3162 & 1870 & 2997 & 2722 & 299 & 95.5 $^{+4.5}_{-7.4}\%$ & 21\\
\citet{2021arXiv211004296H} & 1336 & 342 & 1231 & 1162 & 264 & 82.9 \pm 6.6$\%$ & 21\\
\citet{2021ApJ...923..129J} & 601 & 0 & 592 & 560 & 171 & 95.3$^{+4.7}_{-7.2}\%$ & 18\\
\citet{2021ApJ...917...23K} & 1852 & 274 & 1721 & 1631 & 428 & 82.6 \pm 7.0$\%$ & 20\\
\meingast & 1223 & 104 & 1176 & 1117 & 314 & 86.7 \pm 5.4$\%$ & 19\\
\citet{2022arXiv220604567M} & 165 & 19 & 160 & 151 & 124 & 92.7 \pm 4.4$\%$ & 13\\
\enddata

\tablecomments{Candidate counts in our analysis. N is the number of candidates from each study, N$_{\mathrm{unique}}$ is the number of candidates unique to that study, N$_{\mathrm{LC}}$ is the number of candidates for which we were able to generate a light curve, N$_{\mathrm{Per}}$ is the number of candidates for which we obtained a valid period measurement, N$_{\mathrm{final}}$ is the number of candidates that pass all selection criteria defined in Section \ref{sec: methods}, True Positive Rate is the percentage of stars from each study that are rotationally consistent with $\alpha$ Per membership (see Section \ref{sec:true positive} for a detailed analysis), and $G_{\rm limit}$ is the Gaia apparent magnitude limit of each study's reported $\alpha$ Per members.}
\label{table:counts}
\end{deluxetable*}

\section{TESS Rotation Periods} \label{sec: methods}

We began our analysis by combining the lists of $\alpha$ Per candidates into one main list. This gave us a total of 5754 candidate cluster members.  To measure the rotation periods of these stars, we used the full-frame image data from the two-year primary TESS mission \citep{2015JATIS...1a4003R}.  
Using the \texttt{tess-point} software \citep{2020ascl.soft03001B},  we determined that TESS observed candidate $\alpha$ Per members in sectors 5, 6, 14, 15, 16, 17, 18, 19, 20, 24, 25, and 26 of the primary mission, starting in November 2018 for sector 5 and ending in July 2020 for sector 26. However, due to $\alpha$ Per's low galactic latitude, it lies in a crowded region of the galaxy. TESS's 21-arcsecond pixel size and the high stellar surface density in this region can make standard aperture photometry unreliable. We addressed this challenge by using the difference imaging pipeline \citep{2019zndo...3370324B} developed through the Cluster Difference Imaging Photometric Survey (CDIPS; \citealt{2019ApJS..245...13B}). Briefly summarized, this pipeline uses known stellar locations and reference fluxes from Gaia DR2 to extract flux measurements from TESS full frame images. The photometry is performed on difference images that are constructed for each sector, camera, and CCD by subtracting out an astrometrically-aligned and velocity aberration-corrected reference image from each target full-frame image.

Our photometric reduction yielded 6330 light curves for 5226 candidate cluster members, with more than one light curve per star meaning that the candidate was observed in more than one TESS sector. We were unable to generate light curves for 528 stars in our sample.  The main reason for the difference is that the TESS primary mission was not an all-sky survey: 401 of the missing stars did not fall on the TESS CCDs.  The remaining stars for which we did not generate light curves had null $G_{\rm BP}$ or $G_{\rm RP}$ magnitudes in Gaia DR2, which are necessary for our pipeline to compute a given star's reference brightness in the TESS bandpass.  The stars with null $G_{\rm BP}$ or $G_{\rm RP}$ values tend to be faint stars with a mean $G$ of 18.5 mags.
Our analysis used only the smallest aperture size available, with a radius of 1 TESS pixel, to minimize crowding.  We used the PCA co-trended light curves, following the considerations discussed in Appendix~B of \citet{2021AJ....162..197B}.

After generating the light curves we performed two cleaning steps. First, each light curve was median-normalized. If more than one sector of TESS data was available for a star, we then stitched each sector of TESS data together to make one light curve containing all available TESS data. We then masked 0.7 days at the beginning and end of each spacecraft orbit to remove edge effects that are often present in TESS data.

We used two different methods to search for periodicity in each light curve: Phase Dispersion Minimization \citep{1978ApJ...224..953S} and Generalized Lomb-Scargle \citep{2009A&A...496..577Z}, as implemented in \texttt{astrobase} \citep{wbhatti_astrobase}. We also computed the autocorrelation function (ACF) for each light curve, but did not include it in our analysis as we found that the PDM and GLS methods produce more consistent and robust results for our data. In each light curve, we searched for periods between 0.1 days and 14 days and recorded the first, second, and third most prominent periodogram peaks and the respective powers produced by each method.

Due to the large volume of light curves included in this analysis, we designed automated quality checks to rule out light curves where no significant periodicity is present and only keep light curves that show clear stellar variability. To be included in our sample of stars with rotation consistent with $\alpha$ Per membership, we required each light curve to pass the following series of checks:

\begin{enumerate}
    \item {\it Brightness}: $G < 17$.
    We found that we could not reliably determine the periods of any stars fainter than Gaia DR3 \texttt{phot\_g\_mean\_mag} = 17.
    
    \item {\it Crowding}: There must be no stars that are greater than 1/10th as bright as the target star within 21 arcseconds (1 TESS pixel) of the target star. 
    If multiple stars with rotation signals fall within the aperture used to produce the light curves, it can be hard to tell which star corresponds to which rotation signal. 
    
    \item {\it Internal Consistency}: The rotation period returned by the GLS and PDM methods must be within 5$\%$ of each other.
    If this criterion is met, we adopted the period measured by the PDM method as the period of the light curve. If the periods were not within five percent of each other, we checked to see if either method returned a sub-peak that matched the best period returned by the other method. If so, we flagged this as a possible detection, visually inspected the phase-folded light curve to determine the true period (if present), and assigned to the light curve the period from the method that returned the true period (often the GLS method). If no match was found, we concluded that no significant period was present in the data.
    
    \item {\it Periodogram Strength}: The power of the most prominent periodogram peak had to be less than 0.9 if the PDM period was selected and greater than 0.1 if the GLS period was selected. By default, the GLS periodogram is normalized in the same way as in \citet{1976Ap&SS..39..447L} so that the periodogram power always lies between 0 and 1.
    
    \item {\it SNR}: We defined a signal-to-noise metric as follows.
    \begin{equation}
        \mathrm{SNR = \frac{A_{90-10}}{P2P_{RMS}} \sqrt{N_{cycles}}},
    \end{equation}
    where A$_{90-10}$ quantifies the amplitude of the light curve by measuring the range between the 90th and 10th percentiles in the flux of the light curve. P2P$_\mathrm{{RMS}}$ measures the noise of the light curve by determining the point-to-point variation in the flux and is obtained by calculating the 84th - 50th percentile of the distribution of the sorted residuals from the median value of $\delta$F$_{\rm j}$ = F$_{\rm j}$ - F$_{\rm j+1}$, where j is an epoch index. $\mathrm{N_{cycles}}$ is the number of period cycles present in the data. We required each light curve to have SNR $>$ 6 in order to be included in our sample.
    
    \item {\it External Consistency}: We required each star to fall below the Pleiades slow sequence as defined in \cite{2016AJ....152..113R}.
    To define this cutoff, we fitted a polynomial to the Pleiades color-rotation distribution at $G_{\rm BP}$ - $G_{\rm RP} < 1.4$. Two days were added to the rotation periods calculated by this fit to ensure we did not remove stars too close to the slow sequence. We made no cuts at colors redder than $G_{\rm BP} - G_{\rm RP} = 1.4$ (spectral types later than $\approx$K4V).

\end{enumerate}

Our analysis returned a rotation period for 4835 of the 5226 candidates for which we were able to generate a light curve. Of these 4835 candidates, 855 had at least one light curve that passed our automated quality checks. The number of stars that pass each of our filters is shown in Table \ref{table:period filters}.

\begin{deluxetable*}{cccccccccc}
\tablecaption{The number of candidate cluster members that pass each automated filter}
\tablehead{
\colhead{Paper} & \colhead{N$_{\mathrm{LC}}$} & \colhead{N$_1$} & \colhead{N$_2$} & \colhead{N$_3$} & \colhead{N$_4$} & \colhead{N$_5$} & \colhead{N$_6$} & \colhead{N$_\mathrm{auto}$} & \colhead{N$_\mathrm{final}$}
}

\startdata
\cg & 856 & 668 & 389 & 615 & 719 & 759 & 779 & 237 & 227\\
\citet{2019AJ....158..122K} & 2444 & 1988 & 1137 & 1844 & 2070 & 2171 & 2246 & 702 & 674\\
\lodieu & 2997 & 980 & 638 & 1482 & 1737 & 2524 & 2845 & 312 & 299\\
\citet{2021arXiv211004296H} & 1231 & 798 & 497 & 856 & 976 & 1083 & 1142 & 281 & 264\\
\citet{2021ApJ...923..129J} & 592 & 467 & 270 & 441 & 512 & 529 & 547 & 179 & 171\\
\citet{2021ApJ...917...23K} & 1721 & 1190 & 744 & 1187 & 1355 & 1490 & 1583 & 444 & 428\\
\meingast & 1176 & 874 & 528 & 867 & 993 & 1037 & 1083 & 327 & 314\\
\citet{2022arXiv220604567M} & 160 & 160 & 145 & 140 & 153 & 153 & 150 & 126 & 124
\enddata

\tablecomments{The number of candidate cluster members that pass each automated filter described in Section \ref{sec: methods}, with the subscript in each column name corresponding to the filter number. N$_\mathrm{LC}$ is the number of stars for which we were able to generate a light curve, N$_1$ is the number of stars with $\texttt{phot\_g\_mean\_mag}<$17, N$_2$ is the number of stars with no close and bright companions, N$_3$ is the number of stars with a light curve for which the PDM and GLS methods agree on the period, N$_4$ is the number of stars with a light curve that has a prominent power in the periodogram, N$_5$ is the number of stars with a light curve that passes our signal-to-noise metric, N$_6$ is the number of stars with a rotation period that falls below the the slow sequence of the Pleiades (plus two days), N$_\mathrm{auto}$ is the number of stars that pass each automated check, and N$_\mathrm{final}$ is the number of stars that pass all selection criteria.}
\label{table:period filters}
\end{deluxetable*}

As a final quality check on our rotation periods, we manually inspected each light curve and its phase-folded light curve to ensure that periodicity is present, to correct aliases, and to identify any light curves that should be removed from our sample based on morphology that was not captured in the above quality checks. We assigned a letter grade to each light curve:

\begin{itemize}
    \item j — junk. This designation was assigned to a light curve if the light curve was excessively noisy, dominated by scattered light, or if we were not confident in our period determination. We designated 15 light curves as junk.
    \item c — cluster member. This designation was used for objects that were gyrochronally consistent with $\alpha$ Per membership and had positions on a color--absolute magnitude diagram (CAMD) that were consistent with $\alpha$ Per membership, but that had light curves affected by various systemic and physical effects. Such examples included: eclipsing binaries, light curves where multiple rotation periods are present, and light curves with low-amplitude periodicity. We assigned a ‘c’ designation to 61 objects. 
    \item g — gold. This designation was used for objects that showed clean, clear evidence of rotational variability. We consider this to be our highest quality sample and gave this designation to 748 stars.
    \item f — field. These were stars that, due to their position on a CAMD, showed evidence for being field stars. We assigned the ‘f’ designation to 31 stars and removed them from our analysis before continuing. 
\end{itemize}

After our automated and manual quality checks, we obtained high-quality rotation periods for 809 of the 5226 stars in our original candidate list. In the rest of our analysis, whenever we need a sample of stars with clean rotation periods (e.g., when calibrating gyrochronology at the age of $\alpha$ Per), we use only stars with the ‘g’ designation. Otherwise, we use stars with both the ‘g’ and ‘c’ designation, as the stars in the ‘c’ sample are still consistent with being $\alpha$ Per members.

\subsection{Uncertainties on Rotation Periods} \label{sec:uncertainties}

There is no consensus in the literature on the best way to calculate uncertainties for stellar rotation periods derived from a periodogram analysis. One motivating reason to calculate uncertainties on rotation periods to accurately compare the rotation sequences of one cluster to another. If the rotation periods have large uncertainties, it will be difficult to define where the rotation sequence of the cluster truly lies, making analyses such as deriving ages more difficult. Uncertainties are also useful in long-term studies of stellar evolution, where the quality of historical data is incredibly important and lack of uncertainties can make it difficult to understand the veracity of rotation period measurements.

Some methods for calculating uncertainties on rotation periods include:

\begin{itemize}
    \item Examining the extent to which physical effects, such as spot evolution and differential rotation, affect ability to extract an acceptable phase folded light curve from the data \citep{2020A&A...644A..16G}.
    \item Calculating rotation periods for stars with multiple observations and comparing the periods from each observation \citep{2020A&A...635A..43R}.
    \item Using equations from \citet{1986ApJ...302..757H} or \citet{2004A&A...417..557L} \citep[e.g.,][]{2010A&A...520A..15M}.
    \item Computing the ACF and using the standard deviation in the vertex location of each peak in the ACF as the uncertainty \citep{2022ApJ...936..138H}.
    \item Calculating the full-width at half-maximum (FWHM) or half-width at half-maximum (HWHM) of the most prominent peak in the periodogram \citep{2022ApJ...936..138H, 2022arXiv220711063C, 2020ApJ...903...99H}.
    \item Using Gaussian Processes (GPs) to model stellar variability, with rotation period one hyperparameter included in the GP model \citep{2018MNRAS.474.2094A, 2021ApJ...913...70G}. 
\end{itemize}

Although this is not a complete list of methods present in the literature, it does illustrate the large variety of methods that have been used in the literature to calculate uncertainties on stellar rotation periods. One of the most common methods of calculating uncertainties is by estimating either the full-width at half-maximum (FWHM) or half-width at half-maximum (HWHM) of the most prominent peak in the GLS periodogram or most prominent dip in the PDM periodogram. However, the peak width in the periodogram does not depend on the number of observations or the signal-to-noise ratio present in the data, and often does not change with the quality or quantity of data present \citep{2018ApJS..236...16V}. Plus, the FWHM and HWHM generally overestimate the uncertainties in periodogram results. We therefore do not consider the FWHM or HWHM to be an accurate approximation of the uncertainty. 

Instead, we chose to use an empirical approach to calculating uncertainties. For each light curve, we removed the first 20$\%$ of the light curve and recalculated the periodogram. The part of the light curve that was removed was then added back in, the next $20\%$ removed, and the periodogram recalculated. We repeated this procedure over the length of the light curve and took the standard deviation of the best periods found by each of the five periodogram runs to be the uncertainty in the rotation period measurement. The idea behind this method is that for a high signal to noise light curve that shows strong sinusoidal variation, removing $20\%$ of the light curve before calculating the period will have almost no effect on the period that the periodogram finds, resulting in smaller uncertainties. Conversely, removing $20\%$ of the light curve will have a strong effect on noisy light curves with weaker rotation signals, resulting in a larger uncertainty. We settled on removing $20\%$ of the light curve at a time as removing more of the light curve results in unreasonably large uncertainties (or cases where the periodogram fails to detect a period) and removing less results in no significant change to the calculated periods when compared to the period calculated using the full light curve.

To validate this method, we chose five stars from our sample and calculated uncertainties using our method above and with GPs. To calculate uncertainties using GPs, we followed the method defined in \citet{2021ApJ...913...70G} and used \texttt{celerite} \citep{2017AJ....154..220F} as implemented in \texttt{exoplanet} \citep{foreman_mackey_daniel_2021_7191939} to build a GP model consisting of two quasi-periodic terms to capture the stellar rotation and an additional term to capture any remaining variability in the light curve. The results of this test are as follows: 
\begin{itemize}
    \item Gaia DR2 247179456693727744 ($P_{\rm rot} = 0.159$ d)
    \begin{itemize}
        \item Our uncertainty: $3.00 \times 10^{-5}$ d
        \item GP uncertainty: $2.83 \times 10^{-5}$ d
    \end{itemize}
    \item Gaia DR2 249307187793232768 ($P_{\rm rot} = 1.211$ d)
    \begin{itemize}
        \item Our uncertainty: 0.005 d
        \item GP uncertainty: 0.002 d
    \end{itemize}
    \item Gaia DR2 450859866869548544 ($P_{\rm rot} = 2.307$ d)
    \begin{itemize}
        \item Our uncertainty: 0.771 d
        \item GP uncertainty: 0.996 d
    \end{itemize}
    \item Gaia DR2 247787692782638976 ($P_{\rm rot} = 4.042$ d)
    \begin{itemize}
        \item Our uncertainty: 0.570 d
        \item GP uncertainty: 0.229 d
    \end{itemize}
    \item Gaia DR2 441573048067694592 ($P_{\rm rot} = 6.692$ d)
    \begin{itemize}
        \item Our uncertainty: 0.126 d
        \item GP uncertainty: 0.201 d
    \end{itemize}
\end{itemize}
This test shows that our method returns uncertainties similar to those found using GPs, with the added benefit of being more computationally efficient and easier to implement than GPs. For our analysis, we only calculated uncertainties for stars with the ‘g’ and ‘c’ designations as defined above. Although uncertainties on stellar rotation periods have historically not been handled robustly in the literature, software packages such as \texttt{PIPS} \citep{2022MNRAS.514.4489M} are making progress toward easy-to-use and accurate methods for calculating uncertainties.

\begin{figure*}[ht]
    \centering
    \includegraphics[width=\textwidth]{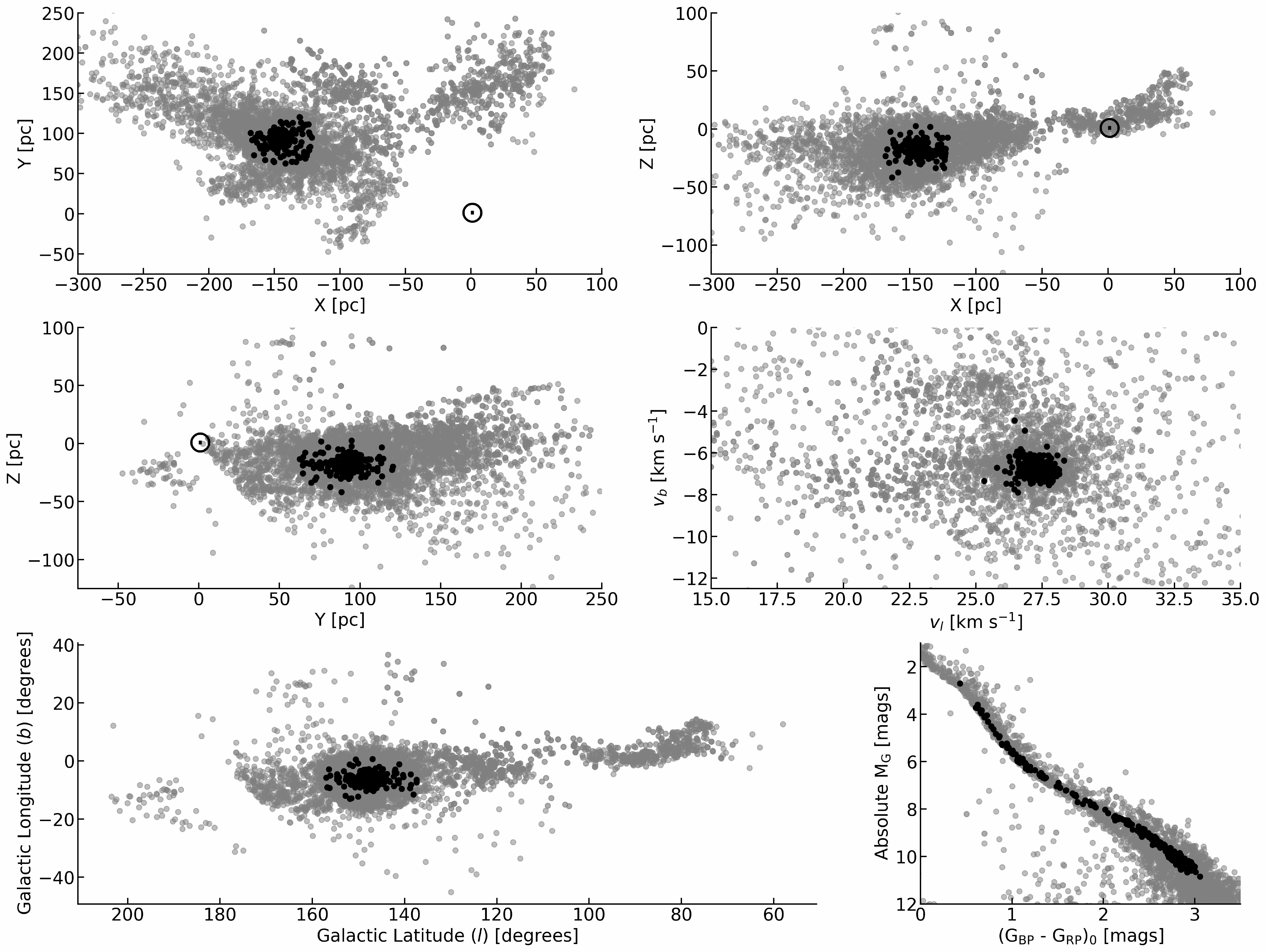}
    \caption{Positions, velocities, and color--absolute magnitude diagram (CAMD) for $\alpha$ Per. The Sun is spatially located at the origin and is represented by the $\odot$ symbol. The grey points represent candidate $\alpha$ Per members from the literature while the black points represent stars that comprise our sample of benchmark rotators. See Section \ref{subsec:gyro sample} for details on how the sample of benchmark rotators was created.}
    \label{fig: fig2}
\end{figure*}

\section{Adopted Age, Reddening, and Effective Temperature Scale} \label{sec:reddening}

Before continuing in our analysis, we corrected for reddening and converted the dereddened $G_{\rm BP} - G_{\rm RP}$ colors to effective temperatures. $\alpha$ Per's low galactic latitude means that it is likely subject to reddening and its large spatial extent may mean that differential reddening should be taken into account.

We corrected for reddening and extinction by using the STILISM dust maps from \citet{2018A&A...616A.132L} and \citet{2017A&A...606A..65C}. The STILISM maps were derived by selecting low-extinction SDSS-APOGEE DR14 red giants and their published empirical extinction coefficients from Gaia and 2MASS. Their extinction coefficients were then compared to their atmospheric parameters to derive extinctions and distance-extinction pairs were inverted to create the map. The STILISM dust maps take as input a star's galactic longitude, galactic latitude, and distance and return the reddening values $E(B-V)$. For each star, we used the Gaia DR2 extinction law presented in \citet{2018A&A...616A..10G} to derive extinction coefficients using $k_m = A_m/A_0$, where $A_0 = 3.1E(B-V)$ and $k_m$ is calculated from Equation (1) of \citet{2018A&A...616A..10G} and the polynomial coefficients presented in their Table 1. This procedure was repeated for each Gaia photometry band (G, BP, and RP). This allowed us to individually correct for each star's extinction and to derive a reddening value for $\alpha$ Per of $E(B-V) = 0.058^{+0.032}_{-0.041}$.

The dereddened Gaia DR2 $G_{\rm BP} - G_{\rm RP}$ values were then converted to effective temperatures by using the calibration from Section A.1 of \citet{2020ApJ...904..140C}. Curtis et al. noted that Gaia DR2 effective temperature values do not account for reddening and are therefore incorrect in regions of substantial reddening, such as in $\alpha$ Per. To remedy this, they built a sample of stars with effective temperatures calculated from three benchmark studies \citep{2012ApJ...757..112B, 2015ApJ...804...64M, 2016ApJS..225...32B} and fit a polynomial to the stars' sequence in color-effective temperature space, allowing them to derive a relation that accurately estimates effective temperatures from dereddened Gaia DR2 photometry with a typical scatter of $\sim$50K. Converting Gaia $G_{\rm BP} - G_{\rm RP}$ colors to effective temperatures has the effect of giving the slow sequence more dynamic range, making it easier to visualize how stellar rotation changes as a function of effective temperature.

\subsection{Empirical Isochrone Age} \label{subsec:isochrone}

We began our age estimation for $\alpha$ Per by first deriving an empirical isochrone-based age for $\alpha$ Per using the method developed by \citet{2020ApJ...903...96G} and implemented in \citet{2022AJ....163..121B}. We began by taking the 5226 stars for which we were able to generate at least one TESS light curve and cleaned our membership list by adopting the quality cuts suggested by \citet{2018A&A...616A..10G}, Appendix B. These cuts were designed to select sources with valid photometry and astrometry while still including binaries. We adopted membership lists for IC 2602 and the Pleiades from \citet{2018A&A...618A..93C} and performed the same filtering on each cluster before continuing.

\begin{figure}[t!] 
    \centering
    \includegraphics[width=0.45\textwidth]{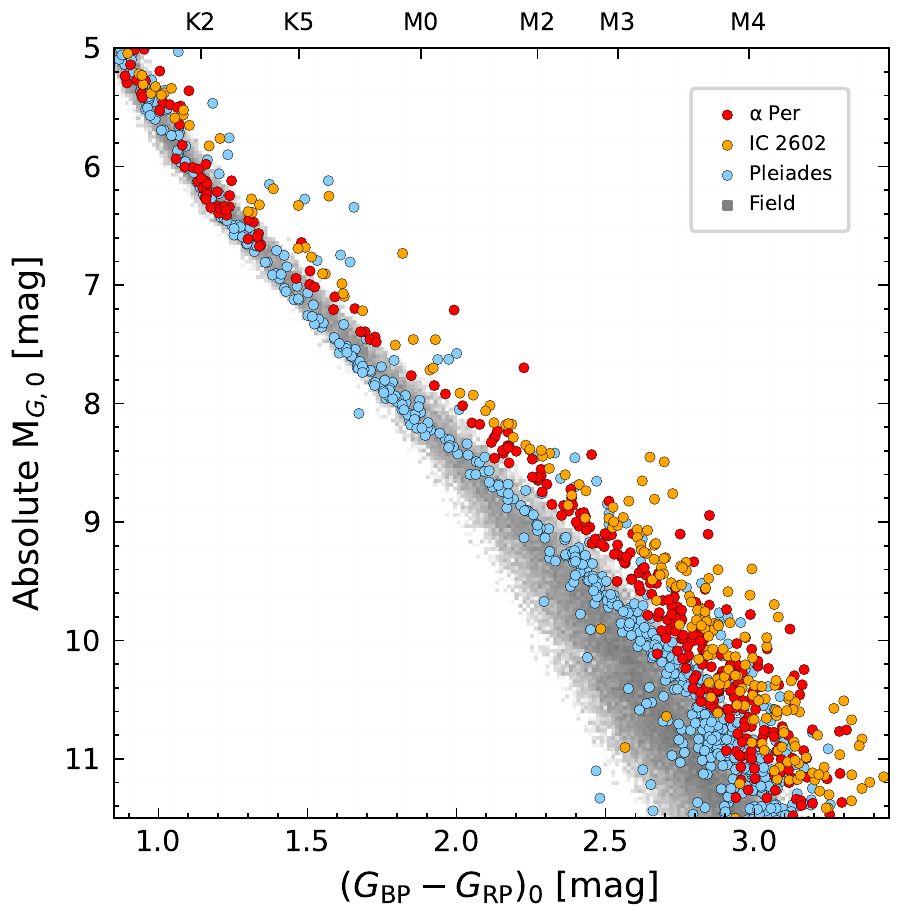}
    \caption{Absolute magnitude as a function of color for K and M dwarfs in $\alpha$ Per, IC 2602, the Pleiades, and the field (grey). The Pleiades is elevated off of the main-sequence, indicating that it is still on the pre-main-sequence, while $\alpha$ Per and IC 2602 are elevated above the Pleiades, indicating that both clusters are younger than the Pleiades.}
    \label{fig:aper_pleaides_ic2602_cmd}
\end{figure}

Figure~\ref{fig:aper_pleaides_ic2602_cmd} shows the results.  The $\alpha$ Per locus falls between the Pleiades and IC 2602, which implies that its age is
intermediate to those two clusters.  Closer inspection reveals that the $\alpha$ Per locus is slightly closer to IC 2602 (38 Myr \citealt{1981A&A....97..235M, 2015ApJ...804..146D, 2015MNRAS.453.2290B, 2018A&A...612A..99R, 2019AJ....158..122K}; 52.5 Myr \citealt{2022A&A...664A..70G}; 46 Myr \citealt{2010MNRAS.409.1002D}) than the Pleiades (112 Myr \citealt{2015ApJ...813..108D}; 127.4 Myr  \citealt{2022A&A...664A..70G}), indicating that $\alpha$ Per is likely closer in age to IC 2602 than it is to the Pleiades.

Before proceeding, we also removed stars with a RUWE value $>$ 1.2 and manually removed any stars that were clear outliers in the CAMD. We then followed the same procedure as in \citet{2022AJ....163..121B}: we binned stars in the CAMD by passing a moving box average over the CAMD in 0.10 mag bins, fitted a spline to the binned values, and generated a piecewise grid of empirical isochrones between the ages of IC 2602 and the Pleiades. To derive a probability distribution function for the age of $\alpha$ Per, we then assumed a Gaussian likelihood that treated the interpolated isochrones as the ``model'' and $\alpha$ Per’s isochrone as the ``data'' \citep[Equation (7) from][]{2020ApJ...903...96G}. The final age and uncertainty values were evaluated using the resulting posterior probability distribution. This method assumed that each cluster evolves toward the ZAMS linearly in time.

\begin{figure*}[t!]
    \centering
    \includegraphics[width=\textwidth]{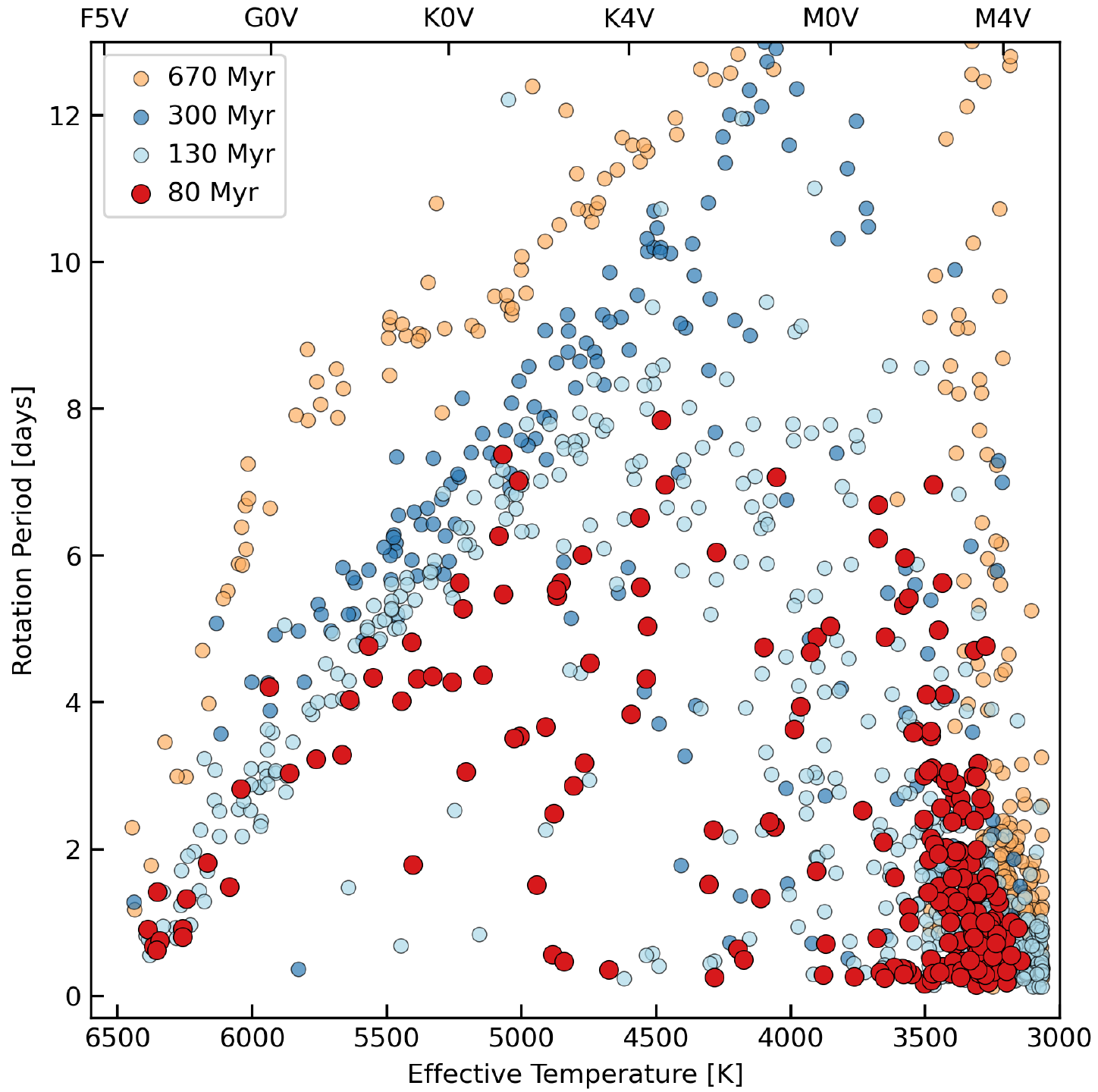}
    \caption{The rotation--effective temperature sequence of $\alpha$ Per (79 Myr) compared against other benchmark clusters. Possible photometric, astrometric, and visual binaries have been removed (see Section \ref{subsec:gyro sample}). This selection against binaries significantly decreases the number of stars on the fast sequence. The 670 Myr sequence is comprised of stars from Praesepe \citep{2021ApJ...921..167R}, the 300 Myr sequence of NGC 3532 \citep{2021A&A...652A..60F} and Group X \citep{2022A&A...657L...3M}, and the 120 Myr sequence of Blanco 1 \citep{2020MNRAS.492.1008G}, the Pleiades \citep{2016AJ....152..113R}, and Pisces-Eridani \citep{2019AJ....158...77C}.  A solar-metallicity, 0.75$M_\odot$ star arrives on the ZAMS at $\approx$80\,Myr with $T_{\rm eff}\approx4700$K.}
    \label{fig:cluter rotation comparison}
\end{figure*}

The resulting empirical pre-main-sequence age for $\alpha$ Per based on its K5V through M3V dwarfs is $\approx$70 Myr, with relative uncertainties of $\sim$15$\%$.  The main uncertainties come from the adopted LDB ages of IC 2602 and the Pleiades, which could skew the inferred empirical isochrone age to anywhere between 60 and 80 Myr.  On a relative scale, $\alpha$ Per seems to be 40-50 Myr younger than the Pleiades, and 20-25 Myr older than IC 2602.  This is consistent with the observation that it is located closer to IC 2602 in the CAMD than to the Pleiades.  To retain an age scale tied to the homogeneous one presented by \cite{2022A&A...664A..70G}, for our adopted $\alpha$ Per age we assume LDB ages of 52.5 Myr for IC 2602 and 127.4 Myr for the Pleiades.  This yields an empirical isochrone age for $\alpha$ Per of $77.5^{+11.9}_{-10.3}$\,Myr.

\section{Gyrochronology at the age of $\alpha$ Per} \label{sec:gyro}

\subsection{Defining A Gyrochronology Sample} \label{subsec:gyro sample}

In order to calibrate gyrochronology, we needed a set of single stars with high-quality rotation period measurements. Due to disk-locking and in rare cases tidal synchronization, binary stars often have faster rotation periods than single stars of the same age and will comprise a significant majority of stars on the fast sequence \citep[e.g.,][]{2016AJ....152..115S,2020MNRAS.492.1008G,2021AJ....162..197B}. We therefore took the following steps to remove potential binary stars from our sample:

\begin{enumerate}
    \item We plotted $\alpha$ Per's sequence in three different color--absolute magnitude diagrams ($M_{\rm G}$ vs $G_{\rm BP} - G_{\rm RP}$, $M_{\rm G}$ vs $G - G_{\rm RP}$, and $M_{\rm G}$ vs $G_{\rm BP} - G$) and manually removed any over-luminous stars in each diagram.
    \item We required Gaia RUWE $<$ 1.2.
    \item We manually removed outliers from a diagram of Gaia DR3 \texttt{radial\_velocity\_error} versus \texttt{phot\_g\_mean\_mag}, which can be indicative of single-lined spectroscopic binarity.
    \item We removed any stars for which the \texttt{non\_single\_star} flag was set in Gaia DR3.
    \item We required the periodogram for each star to not have another peak within 70$\%$ of the height of the main periodogram peak.
\end{enumerate}

We additionally relaxed our crowding requirement (filter 2 from Section \ref{sec: methods}) because we found that the crowding requirement was removing most K and M-dwarfs from our sample. The core of $\alpha$ Per was defined by manually selecting stars that lie in the spatially most concentrated part of the cluster and removing stars that had a velocity dispersion greater than 5 km\,s$^{-1}$ from the median of the manually selected group of stars. The position and velocity of the core is then taken to be the median position and velocity of the remaining stars. We then calculated the velocity difference between each star and the core's median $v_l$ and $v_b$ velocity, and only selected stars within three times the median absolute deviation (MAD) of the \{$v_l, v_b$\} velocity distributions ($\sim$3 km\,s$^{-1}$). The physical distance from each star to the core in XYZ space was also calculated, and we selected only stars within one MAD in spatial distance from the core ($\sim$30 pc). Performing our velocity cuts in two-dimensional \{$v_l, v_b$\} space instead of three-dimensional UVW space allowed us to keep M-dwarfs in our sample, which would otherwise be removed because of Gaia's radial velocity magnitude requirement ($G < 14$). After this cut we were left with 238 stars. Figure~\ref{fig: fig2} shows this sample of stars plotted against all stars in our sample.

\subsection{Stellar Rotation in $\alpha$ Per} \label{subsec:stellar rotation aper}

Figure~\ref{fig:cluter rotation comparison} shows $\alpha$ Per's rotation sequence compared to that of other young, benchmark open clusters: Blanco 1 \citep[$\sim$120 Myr;][]{2020MNRAS.492.1008G}, the Pleiades \citep[$\sim$120 Myr;][]{2016AJ....152..113R}, Pisces-Eridani \citep[$\sim$120 Myr;][]{2019AJ....158...77C}, NGC 3532 \citep[$\sim$300 Myr;][]{2021A&A...652A..60F}, Group X \citep[$\sim$300 Myr;][]{2022A&A...657L...3M}, and Praesepe \citep[$\sim$670 Myr;][]{2021ApJ...921..167R}. Each cluster's list of members was cleaned following the same procedure as in Section \ref{subsec:gyro sample} (minus the check on periodogram powers). As temperature decreases, $\alpha$ Per's rotation periods increase until $\sim$5000K, at which point the slow sequence becomes less defined and the scatter in observed rotation periods increases. At temperatures between $\sim$4500--5000K, the scatter in $\alpha$ Per's rotation periods increases before rotation periods tend to decrease below $\sim$4000K. The transition from increasing to decreasing rotation periods happens at hotter temperatures in $\alpha$ Per than it does for the Pleiades ($\sim$4500K), Group X and NGC 3532 ($\sim$4000K), or Praesepe ($\sim$3500K). The scatter in rotation periods at temperatures less than $\sim$4500K is also smaller in $\alpha$ Per than in the comparison clusters.

The fact that the slow sequence of every comparison cluster lies above that of $\alpha$ Per, the transition from increasing to decreasing rotation periods takes place at hotter temperatures than in the comparison clusters, and the scatter in rotation periods at cool temperatures is less than in the comparison clusters all indicate that $\alpha$ Per is younger than each comparison cluster. This comparison is discussed further in Section \ref{subsec: gyro discussion}.

\begin{figure}[t]
    \centering
    \includegraphics{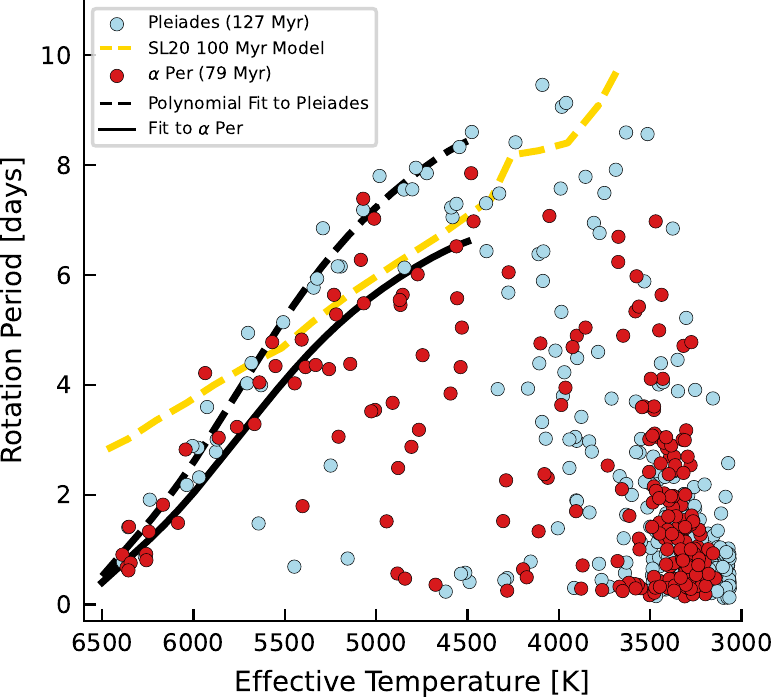}
    \caption{Rotation sequences of the Pleiades and $\alpha$ Per plotted with the 100 Myr gyrochrone from \citet{2020A&A...636A..76S}, a sixth-order polynomial fit to the Pleiades slow sequence, and our model fit to $\alpha$ Per's slow sequence. The model is extended to 4500K even though $\alpha$ Per's slow sequence breaks down at temperatures lower than 5000K.}
    \label{fig: aper_pl_sl_fits}
\end{figure}

In Figure~\ref{fig: aper_pl_sl_fits}, we plot the Pleiades rotation sequence from \citet{2016AJ....152..113R}, our $\alpha$ Per rotation sequence, and the 100 Myr gyrochrone from \citet{2020A&A...636A..76S}. Effective temperatures for the gyrochrones presented in \citet{2020A&A...636A..76S} were interpolated from the given (B-V) colors by using the tables provided by \citet{2013ApJS..208....9P}. The \citet{2020A&A...636A..76S} gyrochrone does not appear to provide a satisfactory fit for the observed rotation sequences of either $\alpha$ Per or the Pleiades. To derive an empirical fit, we manually selected stars on the slow sequence and fitted a sixth-order polynomial to the data. This result is shown in Figure~\ref{fig: aper_pl_sl_fits}.\\\\

\subsection{A Differential Gyrochronology Age for $\alpha$ Per} \label{subsec:gyro age}

We determined a gyrochronology age for $\alpha$ Per by following \citet{2019ApJ...879..100D}, who derived a differential gyrochronology age for the Hyades by comparing the slow sequence of the Hyades to that of Praesepe. We did the same here, except by comparing the slow sequence of $\alpha$ Per to that of the Pleiades. 

Members of the Pleiades were taken from \citet{2016AJ....152..113R}, cleaned following the same procedure defined in Section \ref{subsec:gyro sample}, and their dereddened colors were converted to effective temperature values. As can be seen in Figure \ref{fig:cluter rotation comparison}, the slow sequences for clusters of different ages begin to overlap at temperatures greater than $\sim$6000K. At temperatures less than $\sim$5000K, $\alpha$ Per's slow sequence begins to show significant scatter and loses definition. Gyrochronology at this age therefore seems to be most applicable to stars between $\sim$5000K and $\sim$6000K, so we removed stars outside of this temperature range before proceeding. Any stars that still fall below the slow sequence of $\alpha$ Per or the Pleiades were then manually removed so we could perform a direct comparison between each cluster's slow sequence. 

We used the following equation to derive ages for each star:
\begin{equation} \label{equation:gyro}
    \frac{P_2}{P_1} = \left(\frac{t_2}{t_1}\right)^n
\end{equation}
where $P_2$ and $t_2$ are the rotation period and age of the comparison star, respectively, $P_1$ and $t_1$ are the rotation period and age of the model star, and $n$ is the braking index. In reality this braking index depends on both stellar age and mass \citep{2020ApJ...904..140C}. We therefore could not use the braking index given in \citet{2019ApJ...879..100D} here since their braking index was calibrated for Praesepe ($\sim$670 Myr). If we did use Praesepe's braking index, our resulting differential gyrochronology age would be younger than $\alpha$ Per's true differential gyrochronology age. Instead, we derived a new braking index for $\alpha$ Per by using Equation \ref{equation:gyro} and the Sun. We assumed that the Sun has an age of 4567 Myr \citep{2007leas.book...45C} and a fixed rotation period of 26.09 days \citep{1996ApJ...466..384D}. We also adopted the Gaia DR2 $G_{\rm BP} - G_{\rm RP}$ color of the sun as 0.817 mag from \citet{2019ApJ...879..100D}, which converts to 5789 $\pm$ 50 K using the color-effective temperature relation from \citet{2020ApJ...904..140C}. 

A sixth-order polynomial was then fitted to $\alpha$ Per's slow sequence between 5000 and 6500 K and used to predict the Sun's rotation period at the age of $\alpha$ Per. Fitting a polynomial to temperatures between 5000 and 6500 K ensures we are accurately capturing the shape of $\alpha$ Per's slow sequence. The lithium depletion boundary age given by \citet{2022A&A...664A..70G} for $\alpha$ Per is $79.0^{+1.5}_{-2.3}$ Myr, which is in very good agreement with our empirical isochrone age ($77.5^{+11.9}_{-10.3}$ Myr). If we assumed an age for $\alpha$ Per of 79.0 Myr, this resulted in a predicted solar rotation period of 3.30 days and a braking index of 0.511. If we repeated the same procedure with the Pleiades (assuming an age of $127.4^{+6.3}_{-10}$ Myr from \citealt{2022A&A...664A..70G}), we derived a braking index of 0.546.

Finally, Equation \ref{equation:gyro} was rearranged to derive an age for $\alpha$ Per. We applied this equation to every star in $\alpha$ Per's slow sequence between 5000K and 5700K (15 total stars) to derive a differential gyrochronology age for $\alpha$ Per of 86 $\pm$ 16 Myr. This differential gyrochronology age indicates that $\alpha$ Per is $\sim67 \pm 12\%$ the age of the Pleiades (or $\sim$42 Myr younger than the Pleiades). Our gyro age is consistent with both the LDB age and isochrone age within uncertainties but, given the strong agreement between the LDB age and the isochrone age and the precision of the LDB age, we conclude that the LDB age is the best age determination for $\alpha$ Per.

\section{True Positive Rate of Each Sample} \label{sec:true positive}

Each of the eight papers from which we draw $\alpha$ Per candidates derived their list of candidates using their own method. This begs the question: which method is the most successful at recovering true $\alpha$ Per members? For the purposes of this analysis, we will take the term ``true positive'' to mean the number of stars that are rotationally consistent with cluster membership. There are reasons that a star could be a true cluster member and not have a measurable rotation period, such as by having a pole-on orientation, and those objects will appear as non-members according to our analysis. To assess the true positive rate, from each study we took all candidates for which we were able to generate at least one light curve and only kept candidates that passed the following filters:

\begin{enumerate}
    \item \texttt{phot\_g\_mean\_mag} $<$ 13
    \item There must be no stellar companions greater than 1/10th as bright as the candidate within 21 arcseconds of the candidate.
\end{enumerate}

Since \citet{2022arXiv220604567M} required each of their stars to have a valid Gaia radial velocity (RV) measurement, their sample of cluster members did not extend to magnitudes fainter than $G=13$. As such, we only considered stars brighter than $G=13$ here so that we could directly compare the results of each clustering study. Filter (2) was used when defining our final sample of rotation periods and was applied again here so we could draw an equal comparison between each study and our final sample. After visual inspection of the data, we saw that we are able to confidently detect rotation periods between $0.5 < G_{\rm BP} - G_{\rm RP} < 3.0$. We randomly generated 1000 ranges between $0.5 < G_{\rm BP} - G_{\rm RP} < 3.0$, with a minimum range of 0.1 mags and a maximum range of 2.5 mags. We allowed the low end of each range to vary between $G_{\rm BP} - G_{\rm RP}$ = 0.5 and $G_{\rm BP} - G_{\rm RP}$ = 2.9 and the high end of the range to vary between $G_{\rm BP} - G_{\rm RP}$=0.6 and $G_{\rm BP} - G_{\rm RP}$=3.0, always maintaining a minimum of 0.1 mag between the low and high ends. For each paper, any color range containing less than 5 candidates was thrown away before proceeding. The true positive rate for each range was then calculated as the number of stars from each study with rotation period measurements that passed our quality checks in Section \ref{sec: methods} over the number of candidates from each study. The final true positive rate was taken to be the mean true positive rate over all randomly defined ranges and the uncertainty was the standard deviation in the same sample. There were three main reasons why a star that was identified in a previous study did not pass our quality checks here: the star is too faint (and therefore $P_{\rm rot}$ cannot be measured using TESS), has at least one close and bright companion (and therefore $P_{\rm rot}$ as measured by TESS could be ambiguous), or the star does not have significant stellar rotation. 49.6$\%$ of the stars we consider are fainter than our magnitude limit and 31.0$\%$ of stars have a close and bright companion. These stars were removed from our sample. The remaining stars that failed our automated checks did so because they did not have strong stellar rotation signals — an indication that they are likely to not be cluster members. This procedure gave the following true positive rates (these rates are also reported in Table \ref{table:counts}):

\begin{itemize}
    \item \cite{2018A&A...618A..93C}: 92.6 $\pm$ 7.2$\%$
    \item \cite{2019AJ....158..122K}: 76.7 $\pm$ 7.7$\%$
    \item \cite{2019A&A...628A..66L}: 95.5 $^{+4.5}_{-7.4}\%$
    \item \cite{2021arXiv211004296H}: 82.9 $\pm$ 6.6$\%$
    \item \cite{2021ApJ...923..129J}: 95.3 $^{+4.7}_{-7.2}\%$
    \item \cite{2021ApJ...917...23K}: 82.6 $\pm$ 7.0$\%$
    \item \cite{2021A&A...645A..84M}: 86.7 $\pm$ 5.4$\%$
    \item \cite{2022arXiv220604567M}: 92.7 $\pm$ 4.4$\%$
\end{itemize}

\citet{2018A&A...618A..93C}, \citet{2019A&A...628A..66L}, \citet{2021ApJ...923..129J}, and \cite{2022arXiv220604567M}'s samples all have the highest true-positive rates and are consistent with each other within uncertainties. \cite{2019AJ....158..122K} has the lowest true-positive rate, with 77$\%$ of their sample being composed of true $\alpha$ Per members. 

Out of \cite{2019AJ....158..122K}, \cite{2021ApJ...917...23K}, and \cite{2022arXiv220604567M}, \cite{2022arXiv220604567M} was the only study to require each of their candidate members to have a valid RV measurement (taken from Gaia EDR3). Incorporating Gaia RVs allowed \cite{2022arXiv220604567M} to calculate UVW space positions for each candidate with a precision of 0.1-1.0 km/s, which is comparable to the intrinsic velocity dispersions in open clusters. Their clustering analysis was then performed in six dimensions, whereas the most any other study used was five dimensions. The addition of RV measurements and the extra dimension used in the clustering analysis are likely responsible for the high true positive rate recovered in our work.

\citet{2019A&A...628A..66L}'s clustering analysis is also highly effective, but their membership list was based off of previously reported $\alpha$ Per members, many of whom are spectroscopically confirmed. Since they did not perform a search on Gaia data like the other seven studies here, their true-positive rate is likely biased by selection effects. It should be noted that the true-positive rate quoted above is only for members that Lodieu et al. reported as lying within the tidal radius of $\alpha$ Per. Lodieu et al. performed their analysis in their study using only members within $\alpha$ Per's tidal radius and so we chose to use only those objects here. If we instead perform the true-positive analysis on their full list of candidates that extend to three times $\alpha$ Per's tidal radius, we find that the true-positive rate drops to $90.0 \pm 8.9\%$, showing that Lodieu et al.'s clustering method is still highly effective even at larger distances from $\alpha$ Per's core. Since this analysis only considers stars with $G<13$, we draw no conclusions about contamination at the faint end of each sample.

\begin{deluxetable*}{Ccc} \label{table:full table}
\tablecaption{Rotation Periods and Kinematic Information for 5226 Candidate $\alpha$ Per Members}

\tablehead{
\colhead{Parameter} & \colhead{Example Value} & \colhead{Description}
}

\startdata
\texttt{dr2\_source\_id} & 138206721826797056 & Gaia DR2 source identifier \\ 
\texttt{dr3\_source\_id} & 138206721826797056 & Gaia DR3 source identifier \\ 
\texttt{n\_sectors} & 1 & Number of TESS sectors with CDIPS light curves \\
\texttt{period} & 4.009 & Adopted rotation period (days) \\ 
\texttt{uncertainty} & 1.2$\times10^{-3}$ & Uncertainty on rotation period (days) \\ 
\texttt{pdm\_period} & 4.009 & Period determined using Phase Dispersion Minimization (days) \\ 
\texttt{lsp\_period} & 4.009 & Period determined using Generalized Lomb Scargle (days) \\ 
\texttt{pdm\_power} & 0.181 & Phase Dispersion Minimization periodogram value for best period \\ 
\texttt{lsp\_power} & 0.849 & Generalized Lomb Scargle periodogram value for best period \\ 
\texttt{teff\_curtis20} & 4586 & Effective temperature (K) \\
\texttt{p2p\_rms} & 1580.94 & Peak-to-peak root-mean-square value (ppm) \\ 
\texttt{a\_90\_10} & 38760.76 & Light curve amplitude (ppm) \\ 
\texttt{snr} & 52.46 & Signal-to-noise ratio \\ 
\texttt{ra} & 48.592 & Gaia DR3 right ascension (deg) \\ 
\texttt{dec} & 35.084 & Gaia DR3 declination (deg) \\ 
\texttt{pmra} & 25.353 & Gaia DR3 proper-motion $\mu_{\alpha} \cos \delta$ (mas yr$^{-1}$) \\ 
\texttt{pmdec} & -26.469 & Gaia DR3 proper-motion $\mu_{\alpha}$ (mas yr$^{-1}$) \\ 
\texttt{parallax} & 6.205 & Gaia DR3 parallax (mas) \\ 
\texttt{radial\_velocity} & 2.79 & Gaia DR3 radial velocity (km\,s$^{-1}$) \\ 
\texttt{phot\_bp\_mean\_mag} & 13.267 & Apparent Gaia DR3 G$_\mathrm{{BP}}$ magnitude \\ 
\texttt{phot\_rp\_mean\_mag} & 11.922 & Apparent Gaia DR3 G$_\mathrm{{RP}}$ magnitude \\ 
\texttt{phot\_g\_mean\_mag} & 12.667 & Apparent Gaia DR3 G magnitude \\ 
\texttt{phot\_bp\_mean\_mag\_0} & 7.091 & Gaia DR3 G$_\mathrm{{BP}}$ magnitude (absolute, extinction-corrected) \\ 
\texttt{phot\_rp\_mean\_mag\_0} & 5.803 & Gaia DR3 G$_\mathrm{{RP}}$ magnitude (absolute, extinction-corrected) \\ 
\texttt{phot\_g\_mean\_mag\_0} & 6.523 & Gaia DR3 G magnitude (absolute, extinction-corrected) \\ 
\texttt{X} & -135.53 & Galactocentric X position (pc) \\ 
\texttt{Y} & 68.01 & Galactocentric Y position (pc) \\ 
\texttt{Z} & -52.47 & Galactocentric Z position (pc) \\ 
\texttt{U} & -23.87 & Galactocentric U velocity (km\,s$^{-1}$, corrected for LSR) \\ 
\texttt{V} & -36.27 & Galactocentric V velocity (km\,s$^{-1}$, corrected for LSR) \\ 
\texttt{W} & -13.96 & Galactocentric W velocity (km\,s$^{-1}$, corrected for LSR) \\ 
\texttt{v\_l} & 27.186 & Tangential $l$ velocity (with $\cos b$ correction applied) (km\,s$^{-1}$) \\ 
\texttt{v\_b} & -6.195 & Tangential $b$ velocity (km\,s$^{-1}$) \\ 
\texttt{in\_meingast} & True & Star in \meingast \\ 
\texttt{in\_kerr} & False & Star in \citet{2021ApJ...917...23K} \\
\texttt{in\_kc} & False & Star in \citet{2019AJ....158..122K} \\
\texttt{in\_cg} & False & Star in \cg \\
\texttt{in\_jaehnig} & False & Star in \citet{2021ApJ...923..129J}\\
\texttt{in\_lodieu} & False & Star in \lodieu \\
\texttt{in\_moranta} & False & Star in \citet{2022arXiv220604567M}\\
\texttt{in\_heyl} & False & Star in \citet{2021arXiv211004296H}\\
\texttt{manual\_check} & g & Results of manual check on phase-folded light curve\\
\texttt{flag\_quality\_period} & True & Flag indicating if star has a high quality rotation period detection \\
\texttt{flag\_benchmark\_period} & True & Flag indicating if star is likely a single star \\
\texttt{in\_gyro\_sample} & False & Star appears in sample of stars used to calibrate gyro\\
\texttt{flag\_nbhr\_count} & True & Flag indicating if star has a bright neighbor within a 21 arcsecond radius\\
\enddata

\tablecomments{Table 3 is published in its entirety in a machine-readable format. It contains all 5226 candidates for which we were able to generate a CDIPS light curve. To access the stars used to calibrate gyrochronology in Section \ref{sec:gyro}, use the \texttt{in\_gyro\_sample} flag. Candidates that are rotationally-consistent with $\alpha$ Per membership are flagged using \texttt{flag\_quality\_period}.}

\end{deluxetable*}

\section{Morphology of $\alpha$ Per} \label{sec:morphology}

\begin{figure*}[!tp]
    \centering
    \includegraphics[width=\textwidth]{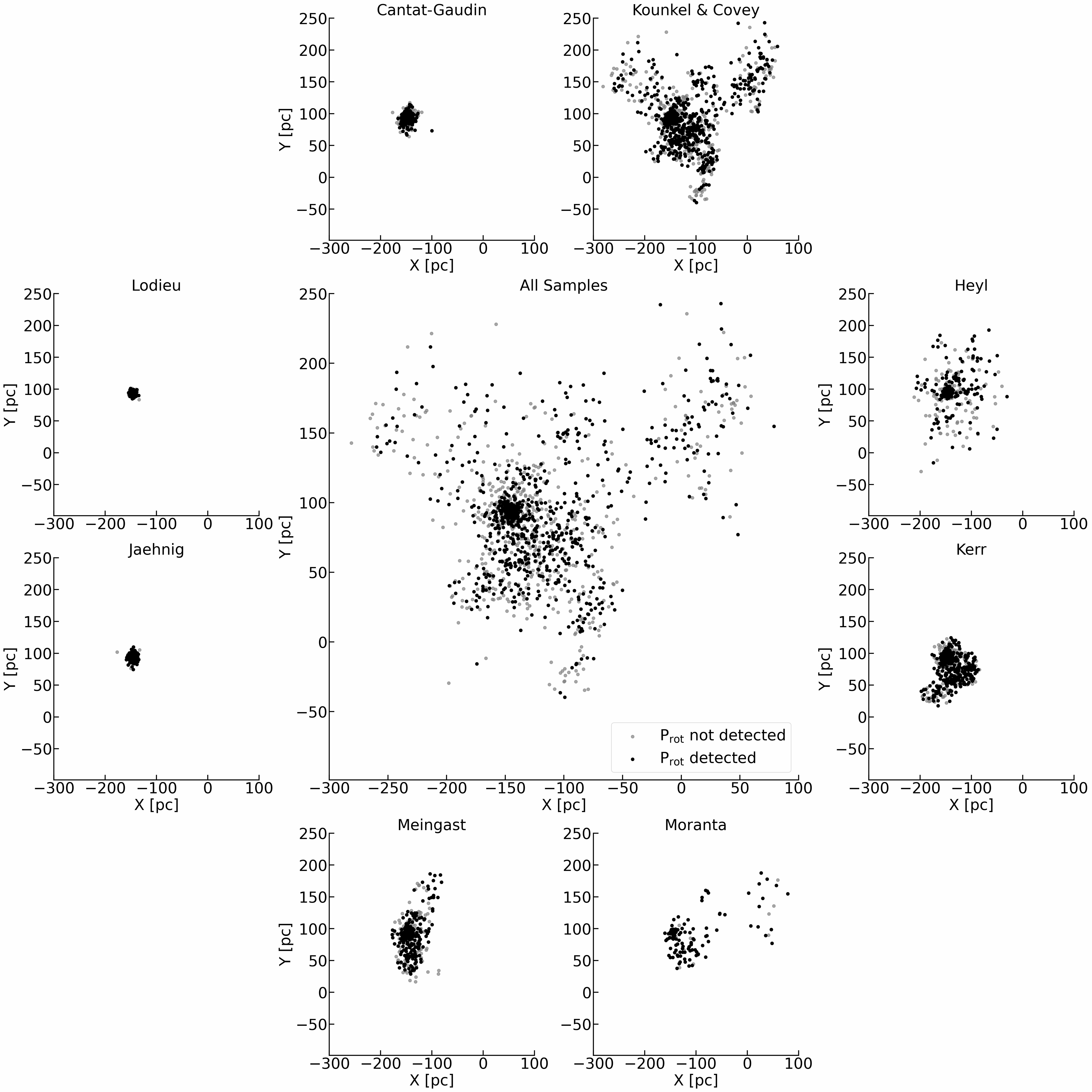}
    \caption{$\alpha$ Per's position in XY space. Galactic rotation is in the $+\hat{Y}$ direction and the galactic center is in the $+\hat{X}$ direction. Each study's selection of $\alpha$ Per members are plotted in the panels around the outside, while the center panel shows all samples combined. Grey points represent cluster members for which rotation periods were expected to be detectable while black points represent stars that are rotationally-consistent with $\alpha$ Per membership.}
    \label{fig:xy positions}
\end{figure*}

\begin{figure*}[!tp]
    \centering
    \includegraphics[width=\textwidth]{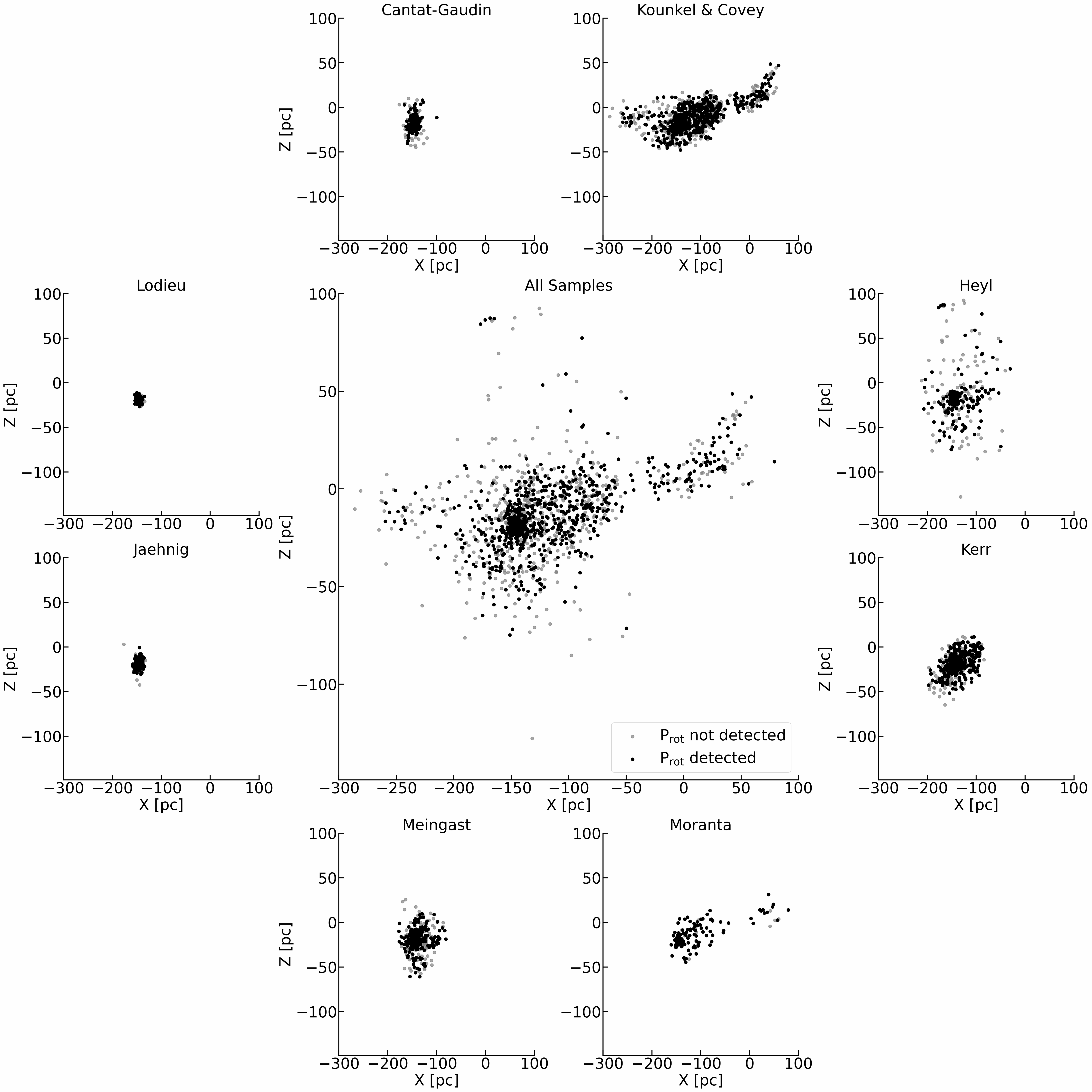}
    \caption{$\alpha$ Per's position in XZ space. The galactic center is in the $+\hat{X}$ direction. Points are as in Figure~\ref{fig:xy positions}.}
    \label{fig:xz positions}
\end{figure*}

\begin{figure*}[!tp]
    \centering
    \includegraphics[width=\textwidth]{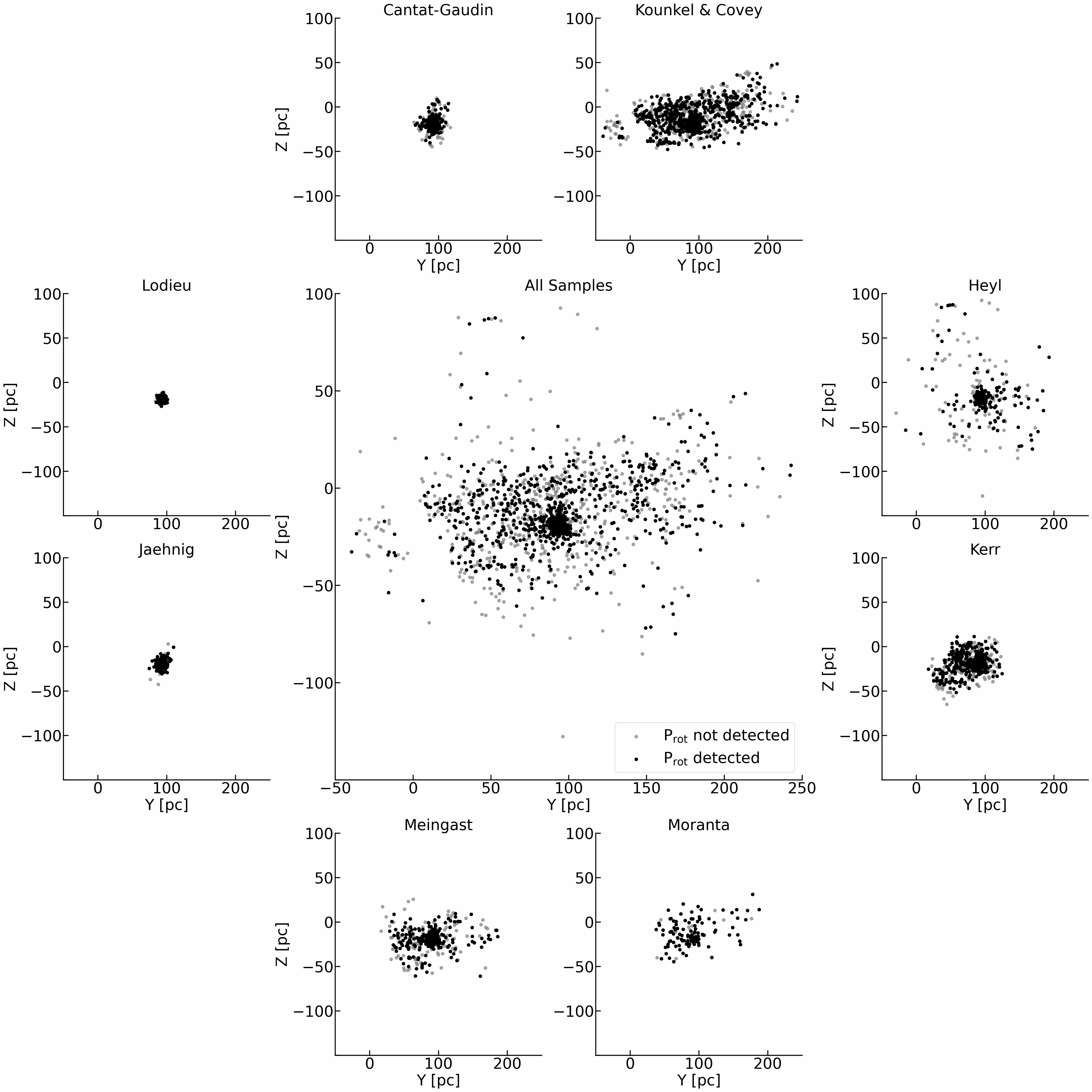}
    \caption{$\alpha$ Per's position in YZ space. Galactic rotation is in the $+\hat{Y}$ direction. Points are as in Figure~\ref{fig:xy positions}.}
    \label{fig:yz positions}
\end{figure*}

\begin{figure*}[tp]
    \centering
    \includegraphics[width=0.95\textwidth]{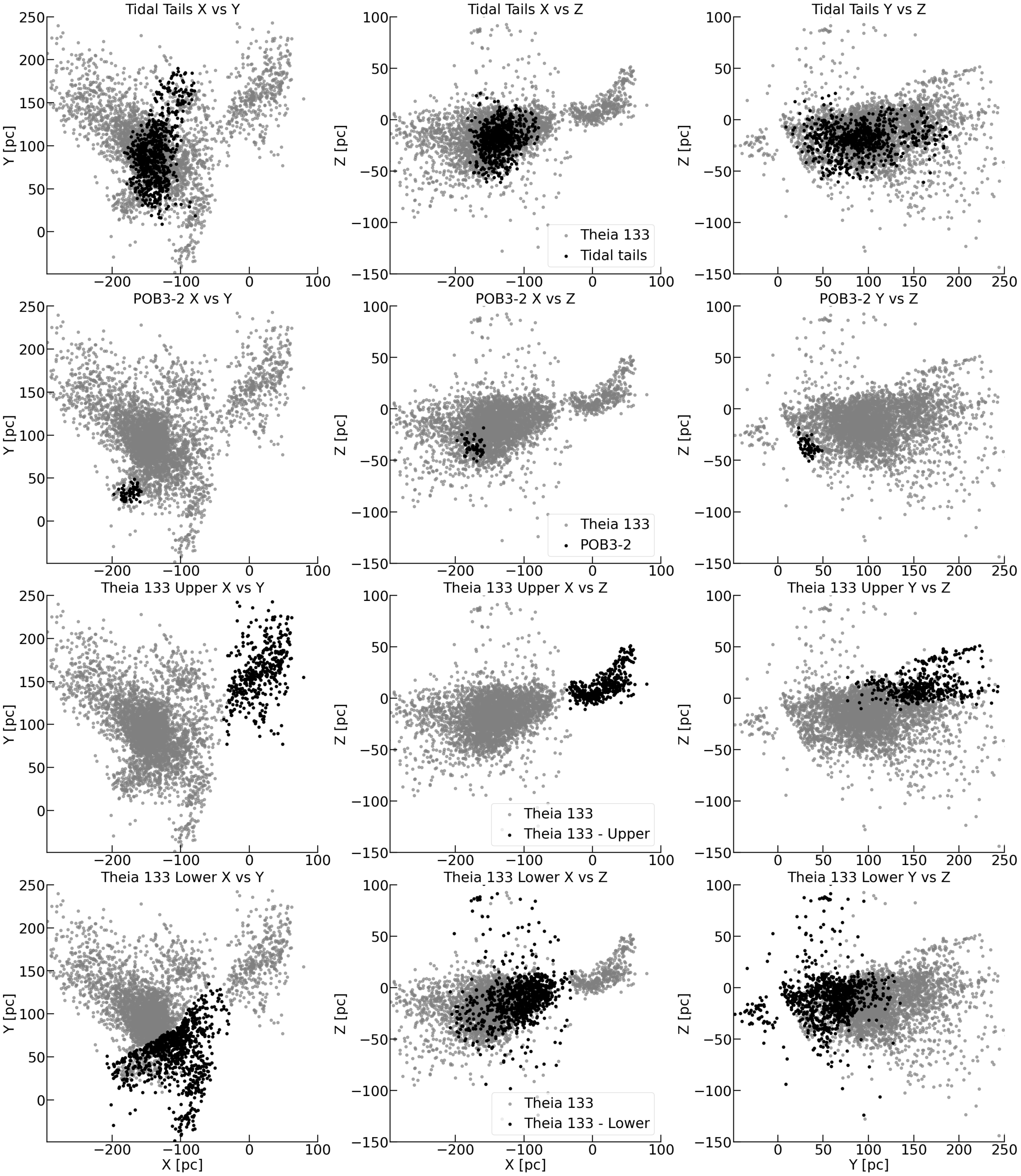}
    \caption{The different regions analyzed to determine $\alpha$ Per's morphology. The grey points indicate all stars in our sample and the black points are the stars that comprise each region. The tidal tails in the top row have a 10 pc radius hole at the center where the core stars have been removed.}
    \label{fig:theia 133 positions}
\end{figure*}

With rotation periods in hand, we turn to investigating the morphology of $\alpha$ Per. Three of the studies included here reported previously undiscovered extensions of $\alpha$ Per. \citet{2021A&A...645A..84M} found tidal tails in $\alpha$ Per extending tens of parsecs away from the core. \citet{2021ApJ...917...23K} designated $\alpha$ Per as POB3-1 and found an additional subcluster $\sim$20$^\circ$ to the southwest, which they designated as POB3-2. The authors noted that POB3-2 lies closer in space to parts of the Greater Taurus region but that there is a branch of low-density stars connecting POB3-2 to $\alpha$ Per. They additionally noted that POB3-2 appears to be kinematically closer to $\alpha$ Per than to Greater Taurus but that they lack the RV coverage to definitively confirm its association with $\alpha$ Per. Finally, \citet{2019AJ....158..122K} reported that $\alpha$ Per is part of part of a much larger group of spatially and kinematically associated stars, which they designated Theia 133. $\alpha$ Per's distribution in XYZ space is plotted in Figures \ref{fig:xy positions}, \ref{fig:xz positions}, and \ref{fig:yz positions}, and the different regions analyzed in this section are visualized in Figure~\ref{fig:theia 133 positions}.

\subsection{Theia 133 from Kounkel $\&$ Covey 2019} \label{subsec:kc morphology}

\paragraph{Theia 133: Upper}
\citet{2019AJ....158..122K} reported that $\alpha$ Per is part of a large spatial structure that they designated as Theia 133. One of the most prominent features of this structure is an additional region extending from $\alpha$ Per's core $\sim$100 pc towards the galactic center and in the direction of galactic rotation. \citet{2022arXiv220604567M} found that one of their newly identified groups, Crius 226, is located in the same region as this extension and appears to be related to $\alpha$ Per, Theia 133, and Theia 209.  \citeauthor{2022arXiv220604567M} also noted that Crius 226 forms a bridge in XY space between Theia 133 and Theia 209. 

\begin{figure*}[ht!]
    \centering
    \includegraphics[width=\textwidth]{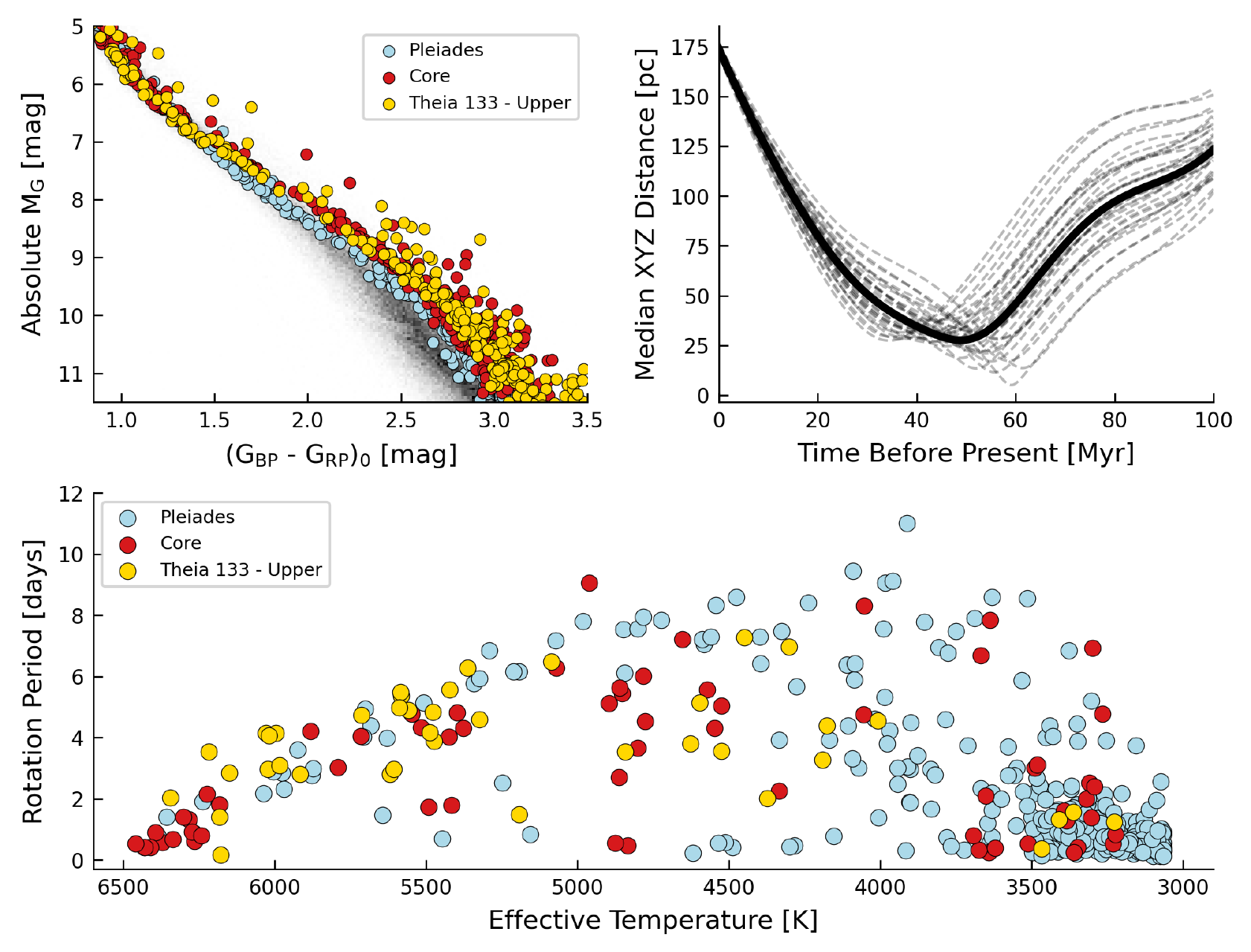}
    \caption{The upper region in Theia 133. Both the rotation and CAMD sequences show that the upper region is a similar age as $\alpha$ Per and the back-integration shows that the median positions of the two regions used to be within $\sim$30 pc of each other. The faint, dashed lines in the upper-right panel represent the uncertainty in the back-integration and are derived from 64 different runs of the back integration, where in each run we change one of the input parameters by one sigma. Of the 418 stars in this region, 85 fall below the Pleiades on the CAMD and have been removed as likely field-star contaminants. Possible binaries have been removed following the procedure defined in Section \ref{subsec:gyro sample}. See Appendix \ref{sec: back int} and Figure \ref{fig:back int potentials} for more discussion on the viability of our back-integrations at young ages.}
    \label{fig:swoop plot}
\end{figure*}

\begin{figure}[t]
    \centering
    \includegraphics[width=0.5\textwidth]{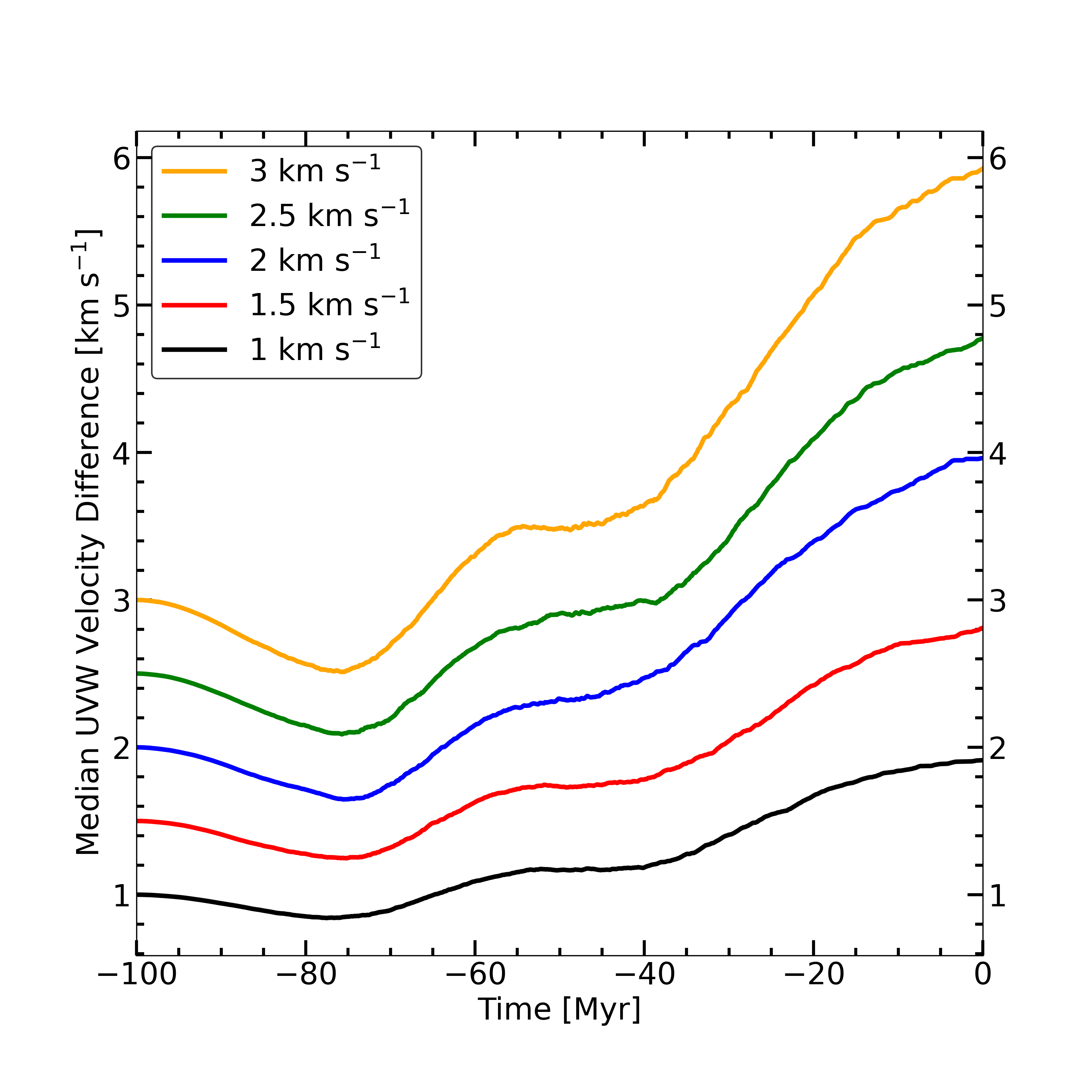}
    \caption{The UVW velocity evolution of test particles over 100 Myr. The initial velocity difference of the test particles almost doubles by 100 Myr, signaling that coeval structures can have large velocity differences even at relatively young ages.}
    \label{fig:uvw}
\end{figure}

Figure~\ref{fig:swoop plot} shows that the rotation sequence for stars in this extended upper region overlaps with $\alpha$ Per's sequence, indicating the two regions are approximately coeval. To further analyze whether these two regions are related, we used \texttt{galpy} \citep{2015ApJS..216...29B} and \texttt{MWPotential2014} to back-integrate both the median positions of each region for 100 Myr. \texttt{galpy} is a galactic dynamics software package that allows the user to model the evolution of orbiting bodies in various gravitational potentials. The potential we used is \texttt{MWPotential2014}, which is a simple and accurate model of the Milky Way's potential and is a combination of individual potentials for each of the Milky Way's bulge, disk, and dark matter halo. The results are shown in the upper-right panel of Figure~\ref{fig:swoop plot}: $\alpha$ Per's core and the upper extension came within $\sim$30 pc of each other 45--50 Myr ago. In other words, these two groups were at least a factor of five closer together in the past, which strongly suggests that they are related to each other. See Appendix \ref{sec: back int} and Figure \ref{fig:back int potentials} for additional discussion on the viability of our back-integrations at young ages. We additionally investigated using LAMOST metallicity measurements to constrain the metallicity of this region but found that LAMOST coverage in this region is sparse and so exclude LAMOST measurements from this part of the analysis.

As an additional check on the upper region, we plotted $\alpha$ Per, the Pleiades, and the upper region on a CAMD in the upper-left panel of Figure~\ref{fig:swoop plot}. $\alpha$ Per is younger than the Pleiades so has more stars on the pre-main-sequence, meaning that it lies above the Pleiades on the CAMD. The upper region lies directly on top of $\alpha$ Per's CAMD sequence, indicating that, like the rotation sequence, the two regions are roughly co-eval.

The median UVW velocity of the upper region differs from that of $\alpha$ Per's core by 4.8 km\,s$^{-1}$. To determine whether coeval groups of stars could have a velocity difference of this magnitude by the age of $\alpha$ Per, we performed the following numerical experiment.  First, to determine a representative starting position and velocity for the core of $\alpha$ Per, we manually selected members of the core using the \texttt{Glueviz} software package \citep{2015ASPC..495..101B, robitaille_thomas_2017_1237692}, calculated their median position and velocity, and then back-integrated for 100 Myr using \texttt{galpy}'s built-in \texttt{MWPotential2014} potential. \texttt{Glueviz} is a software package that allows the user to manually interact with and visualize differences between large data sets. We then defined 1000 test particles in the vicinity of this starting location, and gave them UVW velocities drawn from a uniform distribution ranging between -1 and 1 km\,s$^{-1}$ around the starting UVW velocity of $\alpha$ Per's core.  We also required the total UVW velocity to always fall in a 0.01 km\,s$^{-1}$ range around 1 km\,s$^{-1}$.  Each particle was then integrated forward in time to present day. The velocity difference between $\alpha$ Per's median UVW velocity and the UVW velocity of each test particle was calculated at each step of the integration. We then calculated the median UVW velocity difference between the test particles and $\alpha$ Per at each step to show how the velocity distribution evolved over time. This procedure was repeated for initial UVW velocity differences of 1.5, 2, 2.5, and 3 km\,s$^{-1}$. Figure~\ref{fig:uvw} shows the results. Objects that were born at the location of $\alpha$ Per with an initial velocity difference of 1 km\,s$^{-1}$ from $\alpha$ Per will have a velocity difference of almost 2 km\,s$^{-1}$ after 100 Myr. Similarly, stars with an initial velocity difference of 3 km\,s$^{-1}$ will have a velocity difference of nearly 6 km\,s$^{-1}$ today.

The implication of Figure~\ref{fig:uvw} is that the present-day UVW velocity difference between the core of $\alpha$ Per and the upper-region of Theia 133 is physically plausible given an initial velocity difference of $\sim$2 km\,${\rm s}^{-1}$.  The observed dispersions of young stellar associations easily allows such a difference \citep{2018ApJ...862..138G}. The combination of overlapping rotation and color--absolute magnitude sequences, the small relative velocity difference, and the two regions coming within 30 pc of each other in the past, strongly suggests that the upper region and $\alpha$ Per are indeed related. Once field contaminants have been removed, there are 333 stars in this region (with a total mass of $\sim204M_{\odot}$), with the lowest-mass star having an absolute $M_{\rm G}$ = 15.25 mags \citep[$\sim0.084M_{\odot}$;][]{2013ApJS..208....9P}.

\begin{figure*}[ht!]
    \centering
    \includegraphics[width=\textwidth]{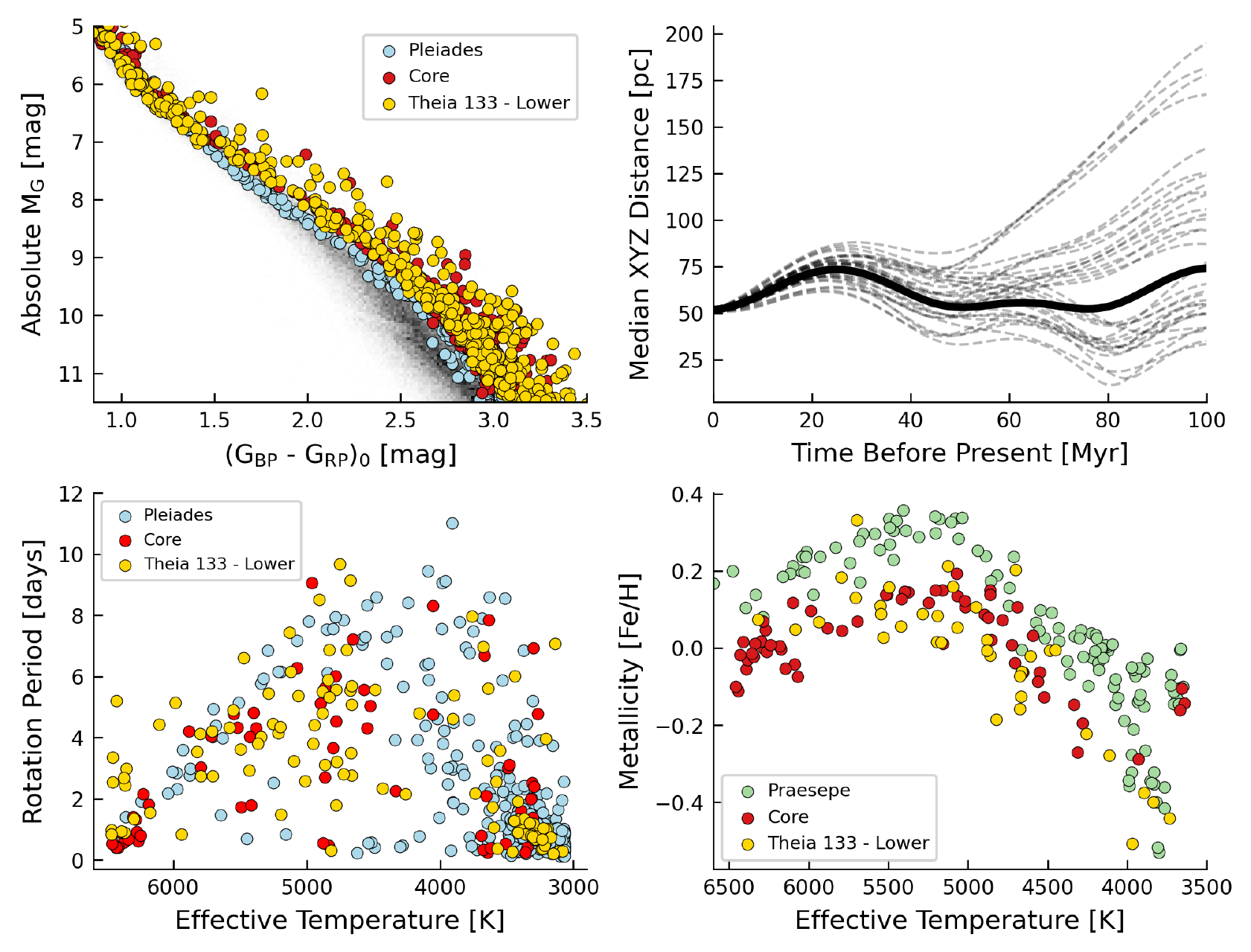}
    \caption{The lower region's color--absolute magnitude, back-integration, rotation, and metallicity sequences. The faint, dashed lines in the upper-right panel represent the uncertainty in the back-integration. All stars in this region that fall below $\alpha$ Per's sequence on the CAMD have been removed as they are likely field-contaminants, leading to a true-positive rate of 74$\%$, which is in agreement with the rotation period derived true-positive rate for \citet{2019AJ....158..122K}. Possible binaries have been removed following the procedure defined in Section \ref{subsec:gyro sample}. }
    \label{fig:lower region}
\end{figure*}

\paragraph{Theia 133: Lower}
\citet{2019AJ....158..122K} reported an additional region that lags behind $\alpha$ Per in XY space, which was also partially recovered by \citet{2021ApJ...917...23K}. To investigate whether this lower region is related to $\alpha$ Per, we performed a similar analysis as for the upper extension to $\alpha$ Per. To define this region's spatial extent, we used \texttt{Glueviz} to manually divide it from the rest of the cluster and removed stars that also overlap with POB3-2, Meingast et al.'s tidal tails, the upper extension, and $\alpha$ Per's core. As can be seen in Figure~\ref{fig:xy positions}, we recovered periods for 209 of the 341 stars that were expected to have rotation periods in this region. These rotation periods are plotted against the rotation sequence of both $\alpha$ Per's core and the Pleiades in Figure~\ref{fig:lower region}. The lower region's slow sequence lies below that of the Pleiades and overlaps with that of $\alpha$ Per. Plus, the transition from increasing rotation periods to decreasing rotation periods occurs at hotter temperatures in this region than in the Pleiades, indicating $\alpha$ Per and this lower region have similar ages. As an additional check on age, we plot the stars that have rotation periods on a CAMD. The lower region's sequence on the CAMD overlaps with $\alpha$ Per's and also indicates they have similar ages.

Groups of stars originating from the same molecular cloud should have the same chemical composition. To explore whether or not this lower region and $\alpha$ Per are co-chemical, we employed LAMOST DR7 metallicities \citep{2012RAA....12.1197C}. After searching LAMOST DR7 LRS data for matches with stars in our sample, we found that LAMOST has measurements for 55 stars in this region. When the lower region's metallicity measurements are plotted over metallicity measurements for $\alpha$ Per's core (Figure~\ref{fig:lower region}), we find that the sequences overlap, indicating that the two regions share similar metallicities. 

As we did for the upper extension, we performed a back-integration for the lower extension. We again used only stars that are rotationally-consistent with $\alpha$ Per membership and back-integrate the median position and velocity of the core and the lower extension. The median XYZ distance between the core and the lower extension over the last 100 Myr is shown in the upper-right panel in Figure~\ref{fig:lower region} and shows that the two regions have remained spatially close over the last 100 Myr, with their distance from each other never increasing or decreasing by more than $\sim25$pc. 

Finally, the lower region's median UVW velocity is separated from the median UVW velocity of $\alpha$ Per's core by only 1.1 km\,s$^{-1}$.  Taken together, the lower region's rotation periods, CAMD, metallicity, back-integration, and velocity provide strongly suggestive evidence that the lower region is related to $\alpha$ Per. This region contains 660 stars for a total mass of $\sim384M_{\odot}$.

\subsection{$\alpha$ Per's Tidal Tails} \label{subsec:tidal tails}

\begin{figure*}[ht!]
    \centering
    \includegraphics[width=\textwidth]{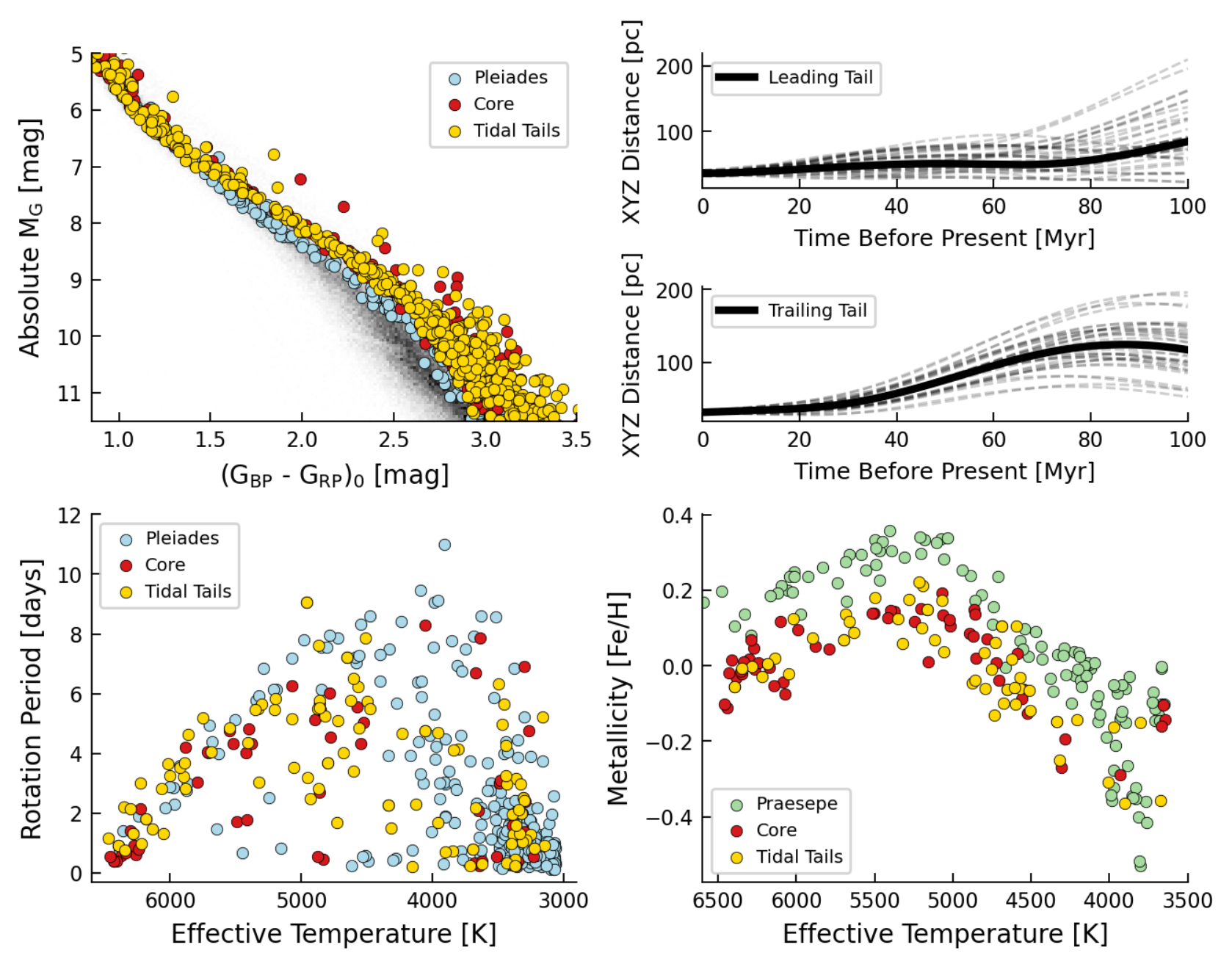}
    \caption{\textit{Upper left}: the Pleiades, $\alpha$ Per's core, and \citet{2021A&A...645A..84M}'s tidal tails on a color--absolute magnitude diagram. Field stars are taken from the Gaia Catalog of Nearby Stars \citep{2021A&A...649A...6G} and are plotted in grey. Tidal tail stars that fall on the Pleiades sequence or below are likely field contaminants and have been removed. \textit{Upper right}: The median XYZ distance between $\alpha$ Per's leading tidal tail and the core over the last 100 Myr. \textit{Middle right}: The median XYZ distance between $\alpha$ Per's trailing tidal tail and the core over the last 100 Myr. The faint, dashed lines in the upper and middle-right panels represent the uncertainty in the back-integration. \textit{Bottom left}: The rotation sequence of the tidal tails compared to that of the core and the Pleiades. Possible binaries have been removed following the procedure defined in Section \ref{subsec:gyro sample}. \textit{Bottom right}: LAMOST metallicity measurements for Praesepe, $\alpha$ Per's core, and the tidal tails.}
    \label{fig:meingast tails}
\end{figure*}

We then turned our attention to the diffuse groups reported by \citet{2021A&A...645A..84M} to lead and lag the core of the cluster.  For brevity, we call these the tidal tails, though their tidal origin has yet to be confirmed. Figure~\ref{fig:xy positions} shows that rotationally-consistent cluster members are clearly present in these regions. To more accurately define stars inside and outside of $\alpha$ Per's tidal tails, we adopted the tidal radius of 10.0 pc from \citet{2021A&A...645A..84M}, which they defined as the radius where the cluster volume density drops below the density in the field. We therefore assumed that any cluster members residing within a 10 pc radius around the center of the core were members of the core and any cluster members residing more than 10 pc from the center of the core were ``corona'' or tidal tail members. We compare $\alpha$ Per's core sequence on a CAMD to that of the tidal tails and the Pleiades in the upper-left panel in Figure~\ref{fig:meingast tails}. The CAMD sequence of the core and the tidal tails overlaps and is elevated above the Pleiades' sequence, indicating that the core and the tidal tails are coeval and younger than the Pleiades. 

The rotation sequence of the tidal tails is plotted in the lower-left panel in Figure~\ref{fig:meingast tails}. Like the CAMD, the rotation sequence of \citet{2021A&A...645A..84M}'s tidal tails overlaps with $\alpha$ Per's slow and fast sequences, indicating that the tails and $\alpha$ Per's core are coeval. Their positioning below the Pleiades and turn over at hotter temperatures than the Pleiades indicates that both the core and the tidal tails are younger than the Pleiades. 

We also compare LAMOST DR7 metallicity measurements for $\alpha$ Per's core and tidal tails in the lower-right panel in Figure~\ref{fig:meingast tails}. The metallicity sequence of the tails and the core overlap, indicating that the stars in each region share similar metallicities. As a final step, we back-integrated the median position and velocity of both the leading and trailing tidal tails. We defined the leading tail as consisting of all stars that are more than 10 pc away from the center of the core in the direction of galactic rotation and the trailing tail as all stars that are more than 10 pc away from the center of the core in the opposite direction of galactic rotation. We find that the median position of each tail was further away from $\alpha$ Per's core in the past --- a finding that is at odds with the idea that the tails formed via evaporation. The upper-right and center-right panels in Figure~\ref{fig:meingast tails} show the median distance between the core and each tail over the last 100 Myr.

These rotation periods, combined with the kinematic and spatial clustering analysis performed by Meingast et al., the overlapping sequences on the CAMD, and similar LAMOST metallicity sequences, suggest that the tidal tails are related to $\alpha$ Per. The back-integration shows the tidal tails moving further away from $\alpha$ Per's core in the past. Possible explanations for this behavior will be discussed in Section \ref{subsec: tidal tails discussion}. 

The binary fraction at different locations in the cluster is a final quantity of dynamical interest. \citet{2020MNRAS.496.5176D} found that the binary fraction in the cores of denser clusters is lower than in the field. If the same holds true here, we would expect $\alpha$ Per's core to have a lower binary fraction than the tails. We estimated the binary fraction in $\alpha$ Per using two different tracers. First, we counted how many stars have a Gaia RUWE $>$ 1.2. \citet{2021ApJ...907L..33S} found that Gaia EDR3 RUWE values even slightly in excess of 1.0 may signify unresolved binaries in Gaia. As such, we made a first comparison of how many stars in each region have RUWE values greater than 1.2. Using our sample of rotationally-consistent cluster members, we find that $11\%$ of stars in the core have RUWE $>1.2$ compared to $15\%$ in the tails. Binaries can also appear as over-luminous stars on a CAMD, so as a second step we cut out all stars in $M_{\rm G}$ vs $G_{\rm BP} - G_{\rm RP}$, $G_{\rm BP} - M_{\rm G}$, and $M_{\rm G} - G_{\rm RP}$ space that lie above $\alpha$ Per's sequence on a CAMD. Combined with the RUWE $>$ 1.2 filter, this gives a binary fraction of $15\%$ for the core and $22\%$ for the tails. We therefore conclude that $\alpha$ Per's core does indeed show evidence of having a lower binary fraction than its tidal tails.  Accounting for the relevant Poissonian uncertainties and selection effects would be an interesting area for future study.

\subsection{POB3-2 from Kerr et al. 2021} \label{subsec:POB3-2}

We broke our analysis of POB3-2 into three parts: refining the velocity difference between $\alpha$ Per and POB3-2, analyzing only stars that \citet{2021ApJ...917...23K} defines as being part of POB3-2, and then additionally considering stars that their clustering analysis identifies as being unclustered members of the POB3 region but that still spatially overlap with POB3-2.

We obtained RV measurements for stars in POB3-2 by querying Gaia DR3 for each of the 74 stars that Kerr defines as comprising POB3-2 and found that 24 of said stars have RV measurements. The RV measurements were then combined with Gaia DR3 right ascension, declination, proper motion, and parallax to derive UVW space velocities for each star. We compared the median UVW space velocity of POB3-2 to that of $\alpha$ Per's core and find that the median velocities of the two regions differ by $\Delta(\mathrm{U,V,W}) = (-6.2, 1.8, 1.1)$ km\,s$^{-1}$ for a total velocity difference of 6.5 km\,s$^{-1}$. 
Referring back to Figure~\ref{fig:uvw}, this velocity difference would imply a primordial velocity difference of $\sim$3.3 km\,s$^{-1}$, assuming that the two structures were initially at the same location.  Given the velocity dispersions of young stellar associations, this is high, but not entirely unreasonable \citep{2018ApJ...862..138G}.

\begin{figure*}[ht!]
    \centering
    \includegraphics[width=\textwidth]{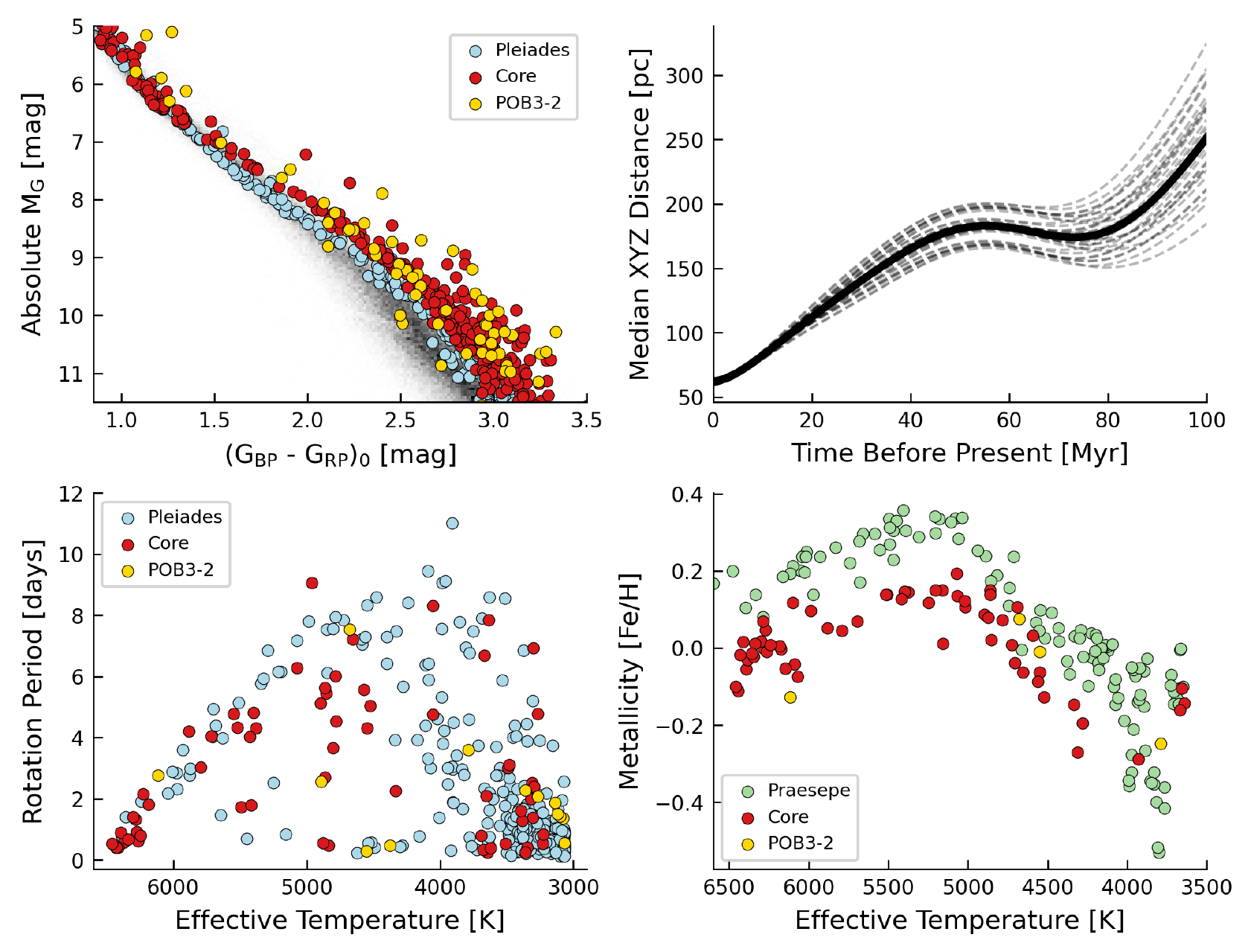}
    \caption{\textit{Upper left}: the Pleiades, $\alpha$ Per's core, and POB3-2 on a color--absolute magnitude diagram. Field stars are taken from the Gaia Catalog of Nearby Stars \citep{2021A&A...649A...6G} and are plotted in grey. \textit{Upper right}: The median XYZ distance between POB3-2 and the core over the last 100 Myr. The faint, dashed lines represent the uncertainty in the back-integration. \textit{Bottom left}: The rotation sequence of POB3-2 compared to that of the core and the Pleiades. Possible binaries have been removed for the core and the Pleiades but not for POB3-2. \textit{Bottom right}: LAMOST metallicity measurements for Praesepe, $\alpha$ Per's core, and POB3-2.}
    \label{fig:pob32 plot}
\end{figure*}

We additionally investigated the rotation sequence, back-integration, and metallicities of POB3-2. Of the 74 stars that Kerr identifies as comprising POB3-2, we recovered rotation periods for only 13 stars, only one of which falls on $\alpha$ Per's slow or fast sequence. In Section \ref{sec: methods}, we recovered rotation periods for 855 out of 5226 total candidates, or $\sim16\%$ of candidates. The recovery rate for POB3-2 is $\sim16\%$ as well. POB3-2's rotation sequence is plotted in the lower-left panel in Figure~\ref{fig:pob32 plot}. Of the 13 stars in POB3-2 with measured rotation periods, only one star passes the possible binary filters defined in Section \ref{subsec:gyro sample} so we plot all 13 stars in the lower-left panel in Figure~\ref{fig:pob32 plot}. The back-integration shows that the median position of the stars in POB3-2 for which we were able to obtain a rotation period only increases in the past relative to $\alpha$ Per's median position. The LAMOST metallicity coverage of POB3-2 was sparse and is included for completeness, though we draw no conclusions from it due to the low number of stars with metallicity measurements in this region. The back-integration and metallicity measurements are shown in the upper-right and lower-right panels in Figure~\ref{fig:pob32 plot}, respectively.

The CAMD sequence for POB3-2 overlaps with $\alpha$ Per's CAMD sequence, indicating that the two regions share a similar age. However, the back-integration, velocity difference, and lack of stars that overlap with $\alpha$ Per's rotation sequence all cast doubt on POB3-2 being related to $\alpha$ Per. If we include the unclustered members of POB3 from Kerr's analysis, the number of stars for which we were able to derive rotation periods rises to 30 and results in three stars appearing to overlap with $\alpha$ Per's slow sequence. We conclude that POB3-2 ($\sim38M_{\odot}$), at least as defined in \citet{2021ApJ...917...23K}, is likely an unrelated population due to its large velocity difference and lack of stars on the slow sequence, but note that there are additional stars present in the area that \citet{2021ApJ...917...23K} does not consider to be members of POB3-2 but that are rotationally-consistent with $\alpha$ Per membership.


\section{Discussion} \label{sec:discussion}

\subsection{White Dwarfs in $\alpha$ Per} \label{subsec:white dwarfs}

White dwarfs can be used to provide an independent age estimate for stellar clusters \citep[e.g.,][]{2002ApJ...574L.155H}. \citet{2015MNRAS.451.4259C} used UKIRT and SuperCOSMOS data to identify 14 white dwarf candidates in $\alpha$ Per but were unable to confirm any of the candidates as bona fide cluster members. \citet{2019A&A...628A..66L} found that two candidates from \citet{2015MNRAS.451.4259C} were in their list of kinematic $\alpha$ Per candidates, but both candidates lie outside of $\alpha$ Per's tidal radius and have total ages older than the cluster's age. \citet{2022ApJ...926L..24M} searched for candidate white dwarfs that might have escaped from $\alpha$ Per in the past and recover two previously-reported candidates (both candidates were also identified by \citet{2019A&A...628A..66L} and \citealt{2021arXiv211004296H}) and discover three new white dwarf candidates at distances greater than 100 pc from the center of the cluster. They confirm that three of their five candidates have cooling ages that are consistent with cluster membership and note that all three are more massive than any other cluster member found using Gaia astrometry. 

The list of $\alpha$ Per candidates defined in Section \ref{sec:sample} can be used to search for additional white dwarf candidates. As such, we cross-match our list of 5226 candidate $\alpha$ Per members with the \citet{2021MNRAS.508.3877G} catalog of white dwarfs in Gaia EDR3 and identify 23 candidate white dwarfs. All 23 of these candidates have probabilities of being a white dwarf in excess of 99$\%$ according to \citet{2021MNRAS.508.3877G}. Two of these white dwarfs are also present in the LAMOST DR8 catalog of white dwarfs: Gaia DR3 173927625524478464 and Gaia DR3 435725089313589376. \citet{2021MNRAS.508.3877G} calculates stellar surface gravity and effective temperature for each star by assuming a pure-H, pure-He, or mixed H/He atmospheric model, and reports parameters for each model. We use \texttt{wdwarfdate} \citep{2022AJ....164...62K}, combined with effective temperature and surface gravity measurements from the Gentile Fusillo et al. catalog or LAMOST DR8 (when applicable) to estimate ages for each white dwarf in our sample. We find that four objects have cooling ages younger than our gyrochronology age for $\alpha$ Per: Gaia DR3 173927625524478464 (cooling age: $29 \pm 1$\, Myr; total age: $2406^{+2990}_{-997}$\, Myr), Gaia DR3 435725089313589376 (cooling age: $47 \pm 2$\, Myr; total age: $266^{+43}_{-36}$\, Myr), Gaia DR3 244003693457188608 (cooling age: $87^{+37}_{-29}$\, Myr; total age: $170^{+36}_{-28}$\, Myr), and Gaia DR3 439597809786357248 (cooling age: $59^{+49}_{-31}$\, Myr; total age: $139^{+51}_{-33}$\, Myr). The total age for each white dwarf exceeds $\alpha$ Per's age. This brings up three possibilities: these white dwarfs are not true $\alpha$ Per members, the white dwarfs are cluster members and the model for their cooling sequence evolution does not account for intrinsic scatter in the population, or our empirical isochrone and gyrochronology ages for $\alpha$ Per are incorrect. Given that the isochrone, gyrochronology, and LDB ages for $\alpha$ Per all agree, we prefer the explanation that these white dwarfs are not true $\alpha$ Per members. Additionally, Gaia DR3 439597809786357248's total age is only 2-$\sigma$ discrepant from $\alpha$ Per's LDB age and empirical isochrone age. Spectroscopic follow-up of Gaia DR3 439597809786357248 would therefore be useful to prove or disprove its association with $\alpha$ Per.\\\\

\subsection{$\alpha$ Per's Metallicity} \label{subsec: metallicity}

\begin{figure}[t]
    \centering
    \includegraphics[width=0.48\textwidth]{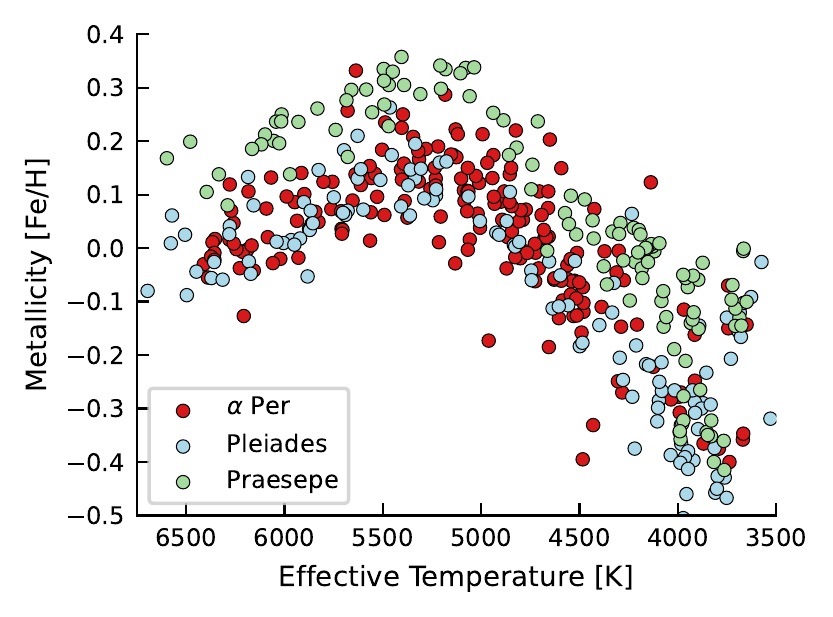}
    \caption{The LAMOST DR7 effective temperature vs. metallicity sequence for Praesepe, $\alpha$ Per, and the Pleiades. $\alpha$ Per's metallicity sequence here is comprised of stars from the core, upper region, lower region, and tails. $\alpha$ Per has a similar metallicity as the Pleiades, and is significantly more metal poor than Praesepe.}
    \label{fig:lamost metallicity}
\end{figure}

As mentioned in Section \ref{sec:aper}, values for $\alpha$ Per's metallicity reported in the literature have ranged from sub-solar \citep[-0.054 $\pm$ 0.046;][]{1990ApJ...351..467B} to super-solar \citep[0.18;][]{2010A&A...514A..81P}. To further explore its metallicity, we cross-match our list of stars that are rotationally-consistent with $\alpha$ Per membership with the LAMOST DR7 LRS Stellar Parameter Catalog of A, F, G and K Stars and LAMOST LRS Catalog of gM, dM, and sdM stars \citep{2022yCat.5156....0L}. The LAMOST metallicity scale is known to systematically vary with effective temperature \citep[e.g.,][]{2022AJ....163..275A}, so we choose to analyze $\alpha$ Per's metallicity by comparing its LAMOST effective temperature vs metallicity sequence to that of the Pleiades and Praesepe. The Hyades were also considered but did not have as complete of coverage in LAMOST as the Pleiades and Praesepe. Figure~\ref{fig:lamost metallicity} shows that Praesepe's sequence falls above that of both $\alpha$ Per and the Pleiades, indicating that Praesepe is the most metal-rich cluster of the three. The sequences of both $\alpha$ Per and the Pleiades overlap, indicating that they have similar metallicities, though $\alpha$ Per's sequence might lie slightly above that of the Pleiades.

In the same spirit as the differential gyro age in Section \ref{sec:gyro}, we now derive a differential estimate of $\alpha$ Per's metallicity by fitting a polynomial to each cluster's LAMOST metallicity sequence. If we compare $\alpha$ Per's measured LAMOST metallicities to those estimated by the fit to Praesepe's sequence, we find that $\alpha$ Per's sequence lies $\sim$0.16 dex below Praesepe's sequence. Assuming a metallicity for Praesepe of $0.21 \pm 0.01$ dex \citep{2020A&A...633A..38D}, we find that $\alpha$ Per has a metallicity of $0.05 \pm 0.03$ dex (median and 1-sigma). For comparison, if we repeat the same procedure to measure the metallicity of the Pleiades, we find a metallicity of $0.03 \pm 0.03$ dex (median and 1-sigma), which is nearly identical to the metallicity reported in \citet{2009AJ....138.1292S}. It therefore appears that $\alpha$ Per's metallicity is marginally super-solar.

\subsection{What is the best method for identifying diffuse stellar structures?} \label{subsec:cluster discussion}

In Section \ref{sec:true positive}, we derived a true-positive rate of $\alpha$ Per members from eight independent studies (Appendix~\ref{sec: clustering descriptions}). There are additional metrics besides the true-positive rate that can be used to assess how well a clustering algorithm performs. For example, a receiver operating characteristic curve can be used to show how the true and false positive rates change at different thresholds and so includes more information than just the true positive rate. Similarly, the false-negative rate is important. \citet{2022arXiv220604567M} for instance had a very high true-positive rate, but missed the larger complex that other studies recovered. Here, we focus on only the true-positive rate and its implications for the effectiveness of each study.

As noted, \citet{2022arXiv220604567M}'s study has the highest true-positive rate of the three studies that used an unsupervised clustering algorithm to perform a blind search on Gaia data, likely due to its inclusion of Gaia RV measurements and its full 6D clustering analysis. \citet{2019AJ....158..122K}, \citet{2021arXiv211004296H}, and \citet{2021A&A...645A..84M} all identify stellar structures beyond $\alpha$ Per's core, with Meingast et al.\ having the highest true-positive rate of the three. It follows that these three studies would have naturally have lower true-positive rates than other studies that focus on identifying cluster cores because more field contaminants will be present further away from cluster cores. 

Choice of input parameters, clustering algorithm settings, and scaling of physical parameters should all be taken into account when performing a clustering analysis and can affect the outcome of the clustering analysis. The first important consideration is what physical parameters should be used for the analysis. For example, should the clustering analysis be performed with sky-projected, galactocentric, or cylindrical positions and velocities, or some combination? In order for galactocentric and cyclindrical velocities to be calculated, each star needs to have a valid RV measurement, many of which are available in Gaia. One drawback of requiring stars to have a Gaia RV measurement is that this requirement will exclude low-mass stars. That being said, not requiring stars to have RV measurements requires a careful treatment of projection effects. \citet{2021ApJ...917...23K} performed their clustering analysis in \{$X, Y, Z, v_l, v_b$\} space with HDBSCAN and find an additional region potentially related to $\alpha$ Per, POB3-2 (see Section \ref{subsec:POB3-2}). However, this region's lack of rotation periods that overlap with $\alpha$ Per's rotation sequence and large UVW velocity difference from $\alpha$ Per's core indicate that it is likely not related. \citet{2021A&A...645A..84M} included an excellent discussion of projection effects as part of their analysis. 

Additionally, these clustering algorithms often require careful thought when setting the parameters used in the analysis, such as the number of stars allowed per cluster and the radius within which to search for stars. Given that changing these input parameters will change the number, size, and distribution of clusters that the algorithm finds, clustering algorithms are somewhat tautological and need verification through other means (as done in this work). The scaling of physical parameters must also be taken into account. The variance in every dimension needs to be similar in relative units because this is all the clustering algorithm sees. Improper feature scaling will result in the parameters that have a broad range of values governing the clustering results. This can be solved by normalizing the input data so that each input parameter contributes proportionally to the clustering algorithm. \citet{2022arXiv220604567M} and \citet{2021ApJ...917...23K} both discuss how to address scaling velocities and positions so they can be clustered together.

\citet{2021A&A...646A.104H} performed a similar analysis on Gaia DR2 data by searching for clusters in Gaia DR2 data with DBSCAN, HDBSCAN, and Gaussian Mixture Models. They find that of the methods they studied, HDBSCAN is the most effective at recovering true-positive cluster members. This is in agreement with our results since HDBSCAN was used in \cite{2022arXiv220604567M} and results in the lowest false-positive rate of all studies included here. However, given that our studies with the lowest and highest true-positive rates both used HDBSCAN, it appears that the choice of physical parameters, clustering settings, and space scaling (as reported above) are just as important of considerations as which clustering algorithm to use.

\cite{2019A&A...628A..66L}, \cite{2021arXiv211004296H}, and \cite{2021ApJ...923..129J} all do not use clustering algorithms such as DBSCAN, HDBSCAN, or UPMASK to identify stellar clusters. Lodieu et al.'s and Jaehnig et al.'s samples have true-positive rates that are consistent with \citet{2022arXiv220604567M} within uncertainties. However, each of these three studies bases their search off of previously-reported positions and/or membership lists for $\alpha$ Per, which likely helps them find true-positive members. \citet{2018A&A...618A..93C} falls into that same category even though it uses a clustering algorithm.

This provides evidence that although clustering algorithms like DBSCAN, HDBSCAN, and UPMASK can be effective tools for recovering stellar clusters, the methods employed by Lodieu et al.\ and Jaehnig et al.\ prove to be equally-effective methods for determining cluster membership that do not rely on unsupervised clustering algorithms. There are also studies not considered here that explore promising clustering techniques, such as combining 3D parameter space with Gaussian Mixture Models and the Mahalanobis distance \citep{2022MNRAS.515.4685D}, using DBSCAN combined with a neural network trained on Gaia photometry \citep{2022A&A...661A.118C}, and using a combination of k-nearest neighbors and Gaussian Mixture Models \citep{2021MNRAS.502.2582A}.

\subsection{What is the true morphology of $\alpha$ Per?} \label{subsec: morphology discussion}

In Section \ref{sec:morphology}, we examined four different proposed extensions to $\alpha$ Per and used isochrone ages, gyrochronology ages, LAMOST metallicities, and back-integration to show that nearby regions of similarly-aged stars are related to $\alpha$ Per. We now turn our attention to discussing what these findings mean for how each region could have formed.

\subsubsection{Theia 133} \label{subsec: theia 133 discussion}

\begin{figure*}[t]
    \centering
    \includegraphics[width=\textwidth]{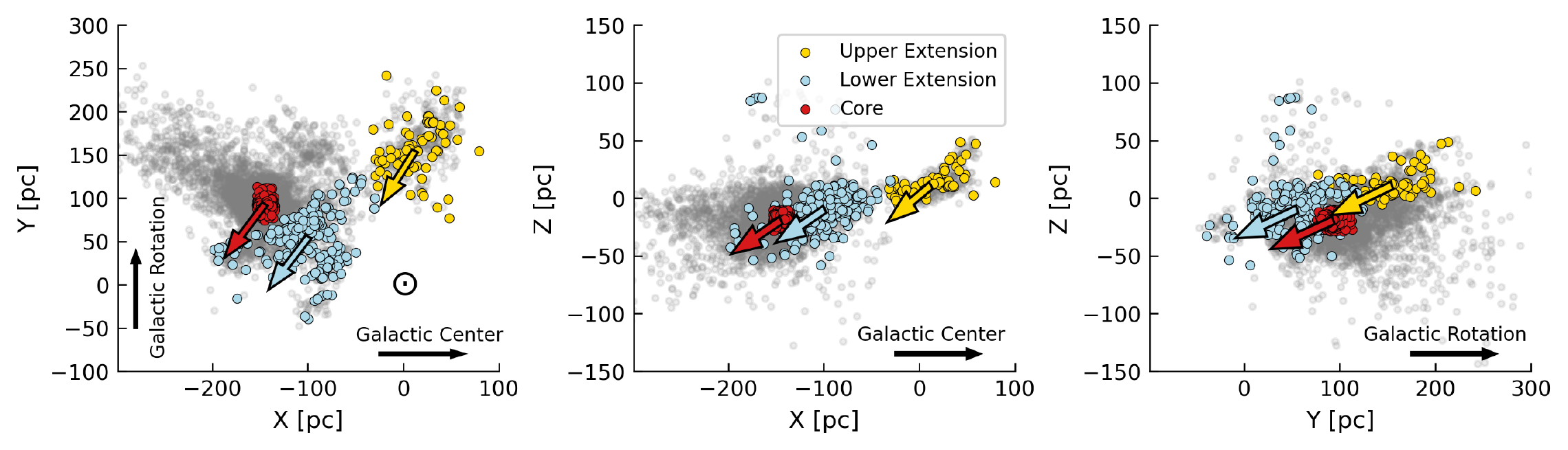}
    \caption{The motions of each region of Theia 133 we analyzed. Grey points represent all stars in our sample for which getting valid rotation periods was possible while colored points are the rotationally-confirmed members in the regions themselves. The arrows represent the median UVW velocities of each region after subtracting the local standard of rest, and their sizes are proportional to the magnitude of each region's median velocity. The Sun is spatially located at the origin and is represented by the $\odot$ symbol.}
    \label{fig:regions motion comparison}
\end{figure*}

In Section \ref{subsec:kc morphology}, we showed that the large, diffuse structures that comprise Theia 133 from \citet{2019AJ....158..122K} do appear to be related to $\alpha$ Per. We now ask: what are these diffuse structures and how did they form?

Our preferred explanation is that $\alpha$ Per is part of a dispersed complex of similarly aged stars and that the upper and lower regions are distinct structures within that complex.  Clusters can inherit hierarchical structure from their molecular clouds \citep[e.g.,][]{2021MNRAS.506.3239G}.  The clouds can have complex shapes, sometimes resembling filaments more than spheres  \citep[e.g.,][]{2018A&A...619A.106G}. Our back-integration from Section \ref{subsec:kc morphology} shows that the core of $\alpha$ Per and the median position of the upper region were within 30 pc of each other in the past. This is well within the typical size scales of molecular filaments \citep{2018ApJ...864..153Z}. If the upper region is composed of stars that evaporated from $\alpha$ Per's core, we would expect the upper region to be made primarily of low mass stars due to mass segregation. However, four of the ten highest mass stars in our sample of stars that are rotationally consistent with $\alpha$ Per membership are located in the upper region. The observed number of high-mass stars in the upper region could be explained if the upper region formed adjacent to $\alpha$ Per, as it would then be its own grouping of stars and not comprised of evaporated stars from $\alpha$ Per's core.

Another possible scenario for forming the upper and lower regions is that the extensions could be part of $\alpha$ Per's tidal tails. The lower extension shares metallicities and both isochronal and gyrochronal ages with $\alpha$ Per, so they are likely related. However, the lower region's back-integration shows that its distance from $\alpha$ Per's core has oscillated by 20 pc over the last 100 Myr while remaining relatively constant. If the lower region is part of $\alpha$ Per's tidal tails, its separation from $\alpha$ Per should decrease in the past, which is not what we see. The separation between the upper region and the core does decrease in the past, but does not come within 30 pc of $\alpha$ Per's core, also indicating that it is likely a separate population. Referring to Figure~\ref{fig:regions motion comparison}, the core, lower extension, and upper extension all appear to be co-moving --- another indication that they are related.

We have shown here that $\alpha$ Per is part of a complex of similarly aged stars but there are other studies that mention even further extensions. \citet{2020ApJ...903...96G} found that the $\mu$ Tau association is very likely related to $\alpha$ Per, Theia 160, Cas-Tau \citep[Cas-Tau was described in][]{2018AJ....156..302D}, and e Tau and u Tau \citep[e and u Tau were originally discovered in][]{2020AJ....159..105L}. \citet{2022arXiv220604567M} notes that they find an additional group, Crius 226, which may be related to $\alpha$ Per, Theia 133, and Theia 209, and forms a bridge in XY space between Theia 133 and Theia 209. With more and more clustering studies being released, it's almost certain that additional clusters will be found that appear to be related to $\alpha$ Per. It would be very worthwhile to repeat the rotation-based analysis presented here on all clusters that appear to be related to $\alpha$ Per in order to gain a better understanding for the true size and morphology of the $\alpha$ Persei complex.

\subsubsection{$\alpha$ Per's Tidal Tails} \label{subsec: tidal tails discussion}

\begin{figure}[t]
    \centering
    \includegraphics[width=0.48\textwidth]{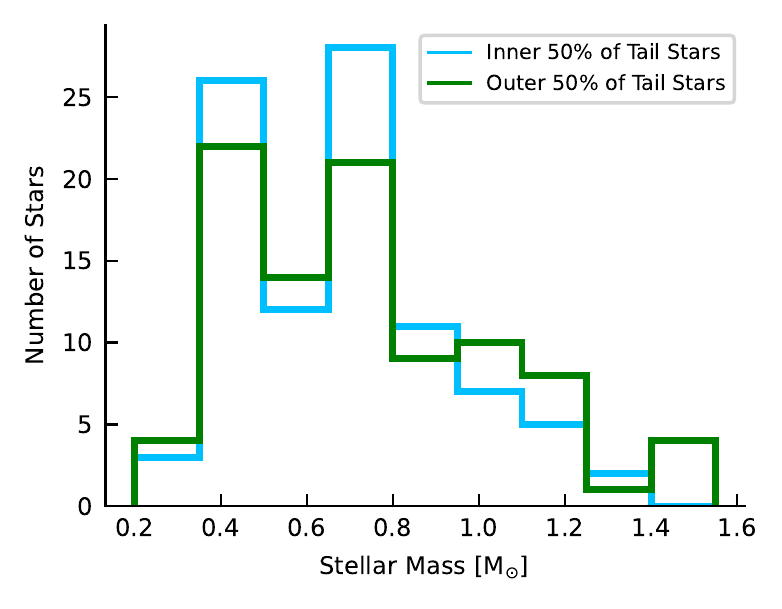}
    \caption{The mass distribution of the 50$\%$ of rotationally-confirmed tidal tail members closest to $\alpha$ Per's core versus the 50$\%$ of members furthest away from $\alpha$ Per's core. The similarity between the two distributions is more consistent with tidal tail formation through gas expulsion than slow evaporation (see Section \ref{subsec: tidal tails discussion}).}
    \label{fig:tidal tails mass}
\end{figure}

Tidal tails can form due to either gas expulsion or dynamical evaporation \citep{2007MNRAS.380.1589B, 2020A&A...640A..84D}. Once star formation in giant molecular clouds forms massive stars, feedback removes most gas and mass from the system, transforming the embedded cluster into an open cluster. During gas expulsion, stars continue to move at their pre-gas expulsion speeds, but since most of the mass has been removed from the system, their outward motions increase. Depending on the star-formation efficiency, clusters can lose as many as 2/3 of their stars during gas expulsion \citep[e.g.,][]{2020A&A...640A..84D}. The remaining stars then revirialize before evaporation of low-mass stars becomes the main source of mass-loss in the cluster. Gas expulsion and evaporation then both contribute to the formation of tidal tails.

However, if the tails formed purely from evaporation, they would be comprised of mostly low-mass stars. Due to mass segregation, low-mass stars should preferentially be located in the outer regions of the cluster. $N$-body simulations predict that tidal tails that form only via gas expulsion should not be mass-segregated, while those that form only via dynamical evaporation should be mass-segregated \citep{2020A&A...640A..85D}. Gas expulsion acts early in a cluster's life and on all stars regardless of mass, while dynamical evaporation acts only on low mass stars and makes them travel at low speeds so they stay closer to the core.

To explore if the mass distribution of the tails, we interpolate stellar masses from our dereddened and corrected Gaia $M_{\rm G}$ values using the tables from \citet{2013ApJS..208....9P}. Figure~\ref{fig:tidal tails mass} shows the 50$\%$ of tidal tail stars that reside closest to $\alpha$ Per's core in XYZ space are comprised of a higher number of low-mass stars and lower number of high-mass stars, while the opposite is true for the 50$\%$ of stars furthest from $\alpha$ Per's core. If the tails formed purely from evaporation, we would expect the least massive stars to be located in the outer reaches of the cluster. Furthermore, our back-integration shows that both the leading and the trailing tails were not closer to the core of the cluster in the past. If the tails formed purely form dynamical evaporation, we would expect the tails to be closer to the cluster in the past. One possible explanation is that the tails themselves are primordial. They are the same age and metallicity as $\alpha$ Per's core so are likely related, but the mass distribution in the tails and their distance from the core in the past mean that the tails could be comprised at least partially of stars that formed in the same vicinity as $\alpha$ Per's core, but not directly a part of it.

The orientation of the tidal tails recovered by \citet{2021A&A...645A..84M} and here also matches expectations from simulations. The Coriolis force causes tidal tails to be elongated in the direction of galactic rotation \citep{2020A&A...640A..85D}. As a cluster moves around the Galaxy, stars that are located closer to the galactic center will experience differential rotation and speed up, forming a leading arm tilted towards the galactic center. Both of these features and shown in Figure~\ref{fig:meingast tails}.

\subsection{How do stars rotate at the age of $\alpha$ Per?} \label{subsec: gyro discussion}

On the pre-main-sequence (PMS), stars contract until their core pressures and temperatures are sufficient to fuse hydrogen, which marks the beginning of the zero-age main sequence (ZAMS).  High-mass stars reach the ZAMS more quickly than low-mass stars.  As a star contracts, its rotation rate generally increases due to conservation of angular momentum.   An exception that can stall this spin-up is disk-locking, which locks the rotation period of the star to that of the inner disk wall \citep{1991ApJ...370L..39K,2018AJ....155..196R}. Once on the ZAMS, angular momentum is carried away by magnetized stellar winds, resulting in spin-down \citep{1967ApJ...148..217W}. For our purposes, there are two factors that affect magnetic braking in the unsaturated regime: the star's mass and rotation. Magnetic braking has a mass dependence that causes low-mass stars with a radiative and convective envelope to spin-down more quickly than high-mass stars \citep{2010ApJ...721..675B, 2015A&A...577A..28J, 2020A&A...635A.170A, 2015ApJ...799L..23M}. This is why the slow sequence has the shape it does --- once on the main-sequence, high mass stars take much longer to spin down than low-mass stars, leading to the trend of increasing rotation period with decreasing mass at a given age.

Internal angular momentum transport may also affect stellar rotation rates \citep{2020A&A...636A..76S}. As PMS stars contract, radiative cores begin to develop and partially replace the already-existing convective cores \citep{2017A&A...599A..49K, 2020A&A...636A..76S}. In the early-PMS, the radiative and convective zones rotate together, making the star roughly a rigid body. However, at the end of the PMS phase, wind braking is at its most efficient and internal angular momentum transport between the radiative and convective zones cannot keep up with the angular momentum loss from wind-braking, leading to differential rotation between the two zones \citep{2020A&A...636A..76S}. This differential rotation persists for the early-main sequence before internal rotational coupling again becomes more efficient than wind-braking. It is the interplay of these two modes of angular momentum transport that is currently the preferred explanation for the stalled spin-down seen in older clusters such as Praesepe and NGC 6811 \citep{2020ApJ...904..140C,2020A&A...636A..76S}. $\alpha$ Per is at an age where wind-braking is expected to be dominant over rotational-coupling at every mass range considered. As such, the observed spin-down seen at effective temperatures $>$5000K in Figure~\ref{fig:cluter rotation comparison} is due to wind-braking, with lower mass stars losing angular momentum more quickly due to the mass-dependence of wind-braking.

$\alpha$ Per's observed rotation periods increase as temperature decreases until temperatures of $\sim$5000K. At this temperature, stars have already converged onto the main-sequence where PMS contraction has stopped and wind-braking is the dominant effect, thereby leading to angular momentum loss and decreasing rotation periods. At temperatures between 4500K and 5000K, the scatter in $\alpha$ Per's observed rotation periods increases and erases the slow sequence seen at hotter temperatures. This temperature range corresponds to a color range of $1.1 < (G_{\rm BP} - G_{\rm RP})_0 < 1.35$. Figure~\ref{fig:aper_pleaides_ic2602_cmd} shows that this color range corresponds to the color at which $\alpha$ Per stars are just beginning to arrive on the ZAMS, meaning that wind-braking and PMS contraction are both affecting stellar rotation. This transition appears to happen at $\sim$4500K for the Pleiades, $\sim$4000K for Group X and NGC 3532, and $\sim$3500K for Praesepe and follows the arrival of each cluster's low-mass stars on the ZAMS. At temperatures below 4500K, rotation periods once again decrease because stars have not yet reached the main-sequence so PMS contraction dominates over magnetic braking, leading to short rotation periods. The apparent scatter in rotation periods at $<$4500K indicates that these stars still contain the rotation information imprinted on them at birth and have yet to lose significant angular momentum.

As stars age, magnetic braking carries away more and more angular momentum, resulting in slower rotation periods at a given mass in older stars. This trend is confirmed here. $\alpha$ Per, the youngest cluster in Figure~\ref{fig:cluter rotation comparison} at 79 Myr, has a slow sequence that lies below that of the other clusters in the figure. Because $\alpha$ Per is younger than the other clusters, magnetic braking has not had a chance to slow stellar rotation as much as it has for the other clusters, leaving $\alpha$ Per's stars to rotate faster at a given mass than in other clusters. The trend mentioned in \citet{2021ApJS..257...46G} that, at a given mass in the saturated regime, the difference in rotation rates between fast and slow rotators increases with age is also recovered in Figure~\ref{fig:cluter rotation comparison}. The difference between the fastest and slowest rotators in Praesepe between 3000K and 3500K is in excess of 20 days, compared to 7.5 days for similar rotators at the age of the Pleiades. This suggests that, at a fixed mass, rapid rotators lose less angular momentum than slow rotators in the saturated regime. $\alpha$ Per's M-dwarfs also appear to have longer rotation periods than stars in the 120 Myr age range. This is because the M-dwarfs at $\alpha$ Per's age have not had as much time to contract as stars in the Pleiades, leading to longer rotation periods.

As noted in several other studies \citep[e.g.,][]{2007ApJ...665L.155M,2021AJ....162..197B, 2022AJ....164..137K}, the fast sequence is mostly comprised of binaries. There are four stars from the Pleiades, three from $\alpha$ Per, and one from Group X that are fast rotators but that survived our binarity filter. One potential explanation for this is disk-locking. In stellar systems with disks, the magnetic field lines from the star can thread the disk, exerting a torque that prevents the star from undergoing PMS contraction. Once the disk dissipates, the star is then free to contract. It is possible that the six stars on the fast sequence in Figure~\ref{fig:cluter rotation comparison} had disks with longer than average lifetimes and so underwent PMS contraction at a later time than other stars in the cluster, resulting in a faster rotation period at a given mass than would otherwise be expected.


\section{Conclusion} \label{sec: conclusion}

TESS rotation periods have been combined with Gaia photometry and astrometry to analyze the $\alpha$ Persei open cluster. Our main conclusions are as follows:

\begin{enumerate}
    \item By $\alpha$ Per's age, single stars hotter than 5000K (K2V, 0.8$M_{\odot}$) have formed a slow sequence. If we assume an LDB age for $\alpha$ Per of $79.0^{+1.5}_{-2.3}$ Myr \citep{2022A&A...664A..70G}, we derive a braking index of 0.51 for stars of this age and effective temperature. At temperatures below 4500K, stars of $\alpha$ Per's age are still on the pre-main-sequence, and at temperatures between 4500K and 5000K, the competing effects of PMS contraction and magnetic spin-down lead to a large variance in observed rotation periods. A 1$M_{\odot}$ star will take $\sim$$40$ Myr to arrive on the ZAMS, while a 0.6$M_{\odot}$ star will take $\sim$$135$ Myr \citep{2016ApJ...823..102C}.  At the age of $\alpha$ Per, a star with mass 0.75$M_\odot$ ($T_{\rm eff}=4700$K) will have just arrived on the ZAMS; it appears some additional time is required for such stars to reach the slow-rotator sequence.
    \item $\alpha$ Per is part of a larger complex of similarly-aged stars.  Three regions appear to be related to the core of $\alpha$ Per: a lower extension ($\sim384M_{\odot}$), upper extension ($\sim204M_{\odot}$), and the tidal tails as reported by \citealt{2021A&A...645A..84M} ($\sim386M_{\odot}$). The lower extension has the same metallicity, isochrone age, and gyrochrone age as $\alpha$ Per, and its separation from $\alpha$ Per's core seems to not have significantly changed over the last 100 Myr. The upper extension similarly shares isochrone and gyrochrone ages with $\alpha$ Per.  Even though the upper extension is currently 175~pc from the core of the cluster, it was five times closer 45--50 Myr ago. The ``tidal tails'' have the expected metallicity, isochrone, and gyrochrone ages, but their back-integration is inconsistent with the expected signature of expansion; they may be primordial.  The lack of mass segregation between the tidal tails and core is a separate line of evidence that supports this possibility.
    \item The most effective clustering analyses leverage Gaia's full 6D position and velocity data to identify bona fide cluster members. True cluster members can be recovered with a high true-positive rate using multiple different clustering algorithms, so the choices of which parameters to use when clustering, the settings adopted in the clustering itself, and how the input data is scaled matter just as much --- if not more --- than the specific algorithm used.
    %
    %
    \item The metallicity of $\alpha$ Per is consistent with the Pleaides within 1-$\sigma$. We use LAMOST DR7 LRS spectra to derive a metallicity for $\alpha$ Per of $0.05 \pm 0.03$ dex.

\end{enumerate}

This analysis makes $\alpha$ Per a new benchmark cluster for studies on the evolution of stellar rotation.  It also provides a new lower anchor for when gyrochronology begins to become applicable.
Beyond stellar rotation, our understanding of $\alpha$ Persei itself is far from complete. With multiple studies now arguing that additional groups are related to the core of $\alpha$ Per, it is clear that more work is needed to understand the full extent and morphology of the complex.   Disentangling how the present-day configuration of $\alpha$ Per was produced by a mix of its primordial structure and dynamical processing remains an exciting prospect for future work.


\begin{acknowledgments}
The authors are grateful to L.~Hillenbrand and M.~Kuhn for helpful discussions and suggestions.
This work was supported by the NASA TESS GI Program (80NSSC22K0298) and the Heising-Simons Foundation 51 Pegasi b Fellowship.

\end{acknowledgments}

%

\facilities{TESS \citep{2015JATIS...1a4003R}, Gaia \citep{2016A&A...595A...1G,2022arXiv220800211G}, LAMOST \citep{2012RAA....12.1197C}}

\software{
Astropy \citep{astropy:2013, astropy:2018, astropy:2022}, astrobase \citep{wbhatti_astrobase}, Matplotlib \citep{Hunter:2007}, pandas \citep{mckinney-proc-scipy-2010, jeff_reback_2022_6702671}, SciPy \citep{2020SciPy-NMeth}, galpy \citep{2015ApJS..216...29B}, Glueviz \citep{2015ASPC..495..101B, robitaille_thomas_2017_1237692}, tess-point \citep{2020ascl.soft03001B}, wdwarfdate \citep{2022AJ....164...62K}, celerite \citep{2017AJ....154..220F}, exoplanet \citep{foreman_mackey_daniel_2021_7191939}
}

\clearpage




\appendix

\section{Descriptions of Each Study's Clustering Analysis} \label{sec: clustering descriptions}

\subsection{Cantat-Gaudin et al. 2018}

\cite{2018A&A...618A..93C} used Gaia DR2 data to derive membership lists and mean parameters for 1229 clusters. The authors started by compiling a list of known clusters and their candidates from previously published catalogs and papers \citep[for $\alpha$ Per:][]{2002A&A...389..871D, 2013A&A...558A..53K}. A cone search centered on the cluster position from the literature was then performed and all stars with \texttt{phot\_g\_mean\_mag} $<$ 18 and a parallax within 0.5 mas of their expected parallax were kept. The authors noted that $\alpha$ Per has a large apparent proper motion dispersion and applied no additional constraints on proper motion before running the cluster through the membership assignment code UPMASK. UPMASK uses k-means clustering to identify small groups of stars in 3D astrometric space ($\mu_{\alpha} \cos \delta$, $\mu_{\delta}$, $\pi$). A “veto” step was then performed where the groups identified by UPMASK were compared to a random distribution and a binary yes/no was returned if the group was more/less concentrated than the random distribution. To determine membership probability, the authors performed the grouping and vetoing step multiple times, each time redrawing new values of $\mu_{\alpha} \cos \delta$, $\mu_{\delta}$, and $\pi$. This procedure was run 10 times and produced a final list of 873 candidate $\alpha$ Per members.

\subsection{Kounkel $\&$ Covey 2019} \label{subsec:KC19}

\cite{2019AJ....158..122K} used Gaia DR2 to derive shapes and ages for clusters, associations, and comoving groups within 1 kpc and $|b|$ $<$ 30 degrees. Their search focused on the galactic mid-plane and extended solar neighborhood between 100 and 1000 pc. Unlike some other papers in our sample \citep{2018A&A...618A..93C, 2021A&A...645A..84M}, \citet{2019AJ....158..122K} did not perform a literature search to get initial positions and/or velocities for stellar clusters. Instead, they applied a series of filters to the full Gaia DR2 sample ($|b|$ $<$ 30 degrees, parallax $>$ 1 mas, and six other quality cuts) to derive a sample of 19.55 million stars. The 5D data set of galactic coordinates $l$ and $b$, parallax, and proper motion from this sample was then run through HDBSCAN. They then iterated over several different parallax cuts to improve sensitivity to associations between 100 pc and 1000 pc from the Sun. The authors noted that the galactic coordinate grid is discontinuous at $l = 0 = 360$ degrees and that some structures that cross this boundary became artificially split. For the structures that became split, the authors performed multiple runs of HDBSCAN for different ranges of $l$ and manually stitched their results together. \citet{2019AJ....158..122K} found 2643 members of $\alpha$ Per (which they designated as Theia 133).

\subsection{Lodieu et al. 2019} \label{subsec:lodieu}

\cite{2019A&A...628A..66L} used a combination of Gaia DR2 data and previously published literature data to provide updated distances, kinematics, and membership lists for $\alpha$ Per, the Pleiades, and Praesepe. The authors combined membership lists from 13 previously published surveys of $\alpha$ Per to make a collection of candidate members, which they crossmatched with Gaia DR2 and kept only those with high-quality astrometry (parallax $>$ 1 mas, RUWE $<$ 1.4). To determine cluster membership, the authors used the kinematic procedure from \cite{1998A&A...331...81P}. Briefly, this procedure involved transforming Gaia parallaxes into distances, calculating the barycenter and space velocity of the cluster, then estimating the expected transverse and radial velocities at the position of each candidate member. Candidates were selected if they had velocities within $\sim4.4\sigma$ of the common cluster motion. Any outliers were then discarded before the mean cluster barycenter and velocity were recalculated. This process was repeated until no outliers remain. The authors calculated the tidal radius of $\alpha$ Per ($\sim$9.5 pc) and reported as their candidates all stars that passed the kinematic procedure outlined above and that were within three times the tidal radius of $\alpha$ Per, for a total of 2069 stars. We note that the authors reported all of their candidates, regardless of distance to the center of $\alpha$ Per, in their list of $\alpha$ Per candidates available online. There are 3162 stars available in this table. In the this analysis, we only used the 517 candidates that Lodieu marked as being within the tidal radius of $\alpha$ Per unless otherwise specified. 

\subsection{Heyl et al. 2021} \label{subsec: heyl}

\cite{2021arXiv211004296H} used Gaia EDR3 to determine past and present cluster members for $\alpha$ Per, NGC 2451A, IC 2391, and IC 2602. The authors started by conducting two cone searches centered on $\alpha$ Per’s position from \cite{2018A&A...616A..10G}. The aim of the two cone searches was to construct a complete sphere of radius 60 pc around the cluster with a narrow cone and a complete hemisphere of radius 90 pc on the nearside of the cluster with a broad cone. In the case of $\alpha$ Per, this selected all objects in Gaia EDR3 within 250 pc of the sun that lie within 28 degrees of $\alpha$ Per on the sky and a second region within 200 pc of the sun and within 45 degrees of the cluster on the sky. The mean cluster position and velocity were then calculated and the sample of candidate stars was defined to be all stars within 10 pc of the cluster center that have proper motions within 5 mas yr$^{-1}$ of the cluster’s median proper motion. Projection effects were not accounted for. An additional analysis was then performed to look for stars that have escaped from $\alpha$ Per. This was done by calculating the velocity of each star relative to $\alpha$ Per and its distance from $\alpha$ Per as a function of time. These values were then compared to the star’s current position, with the reasoning being that each star has to have moved to its current position within the lifetime of $\alpha$ Per. They found 1336 candidate $\alpha$ Per members.

\subsection{Jaehnig et al. 2021} \label{subsec: jaehnig}

\cite{2021ApJ...923..129J} used Extreme Deconvolution Gaussian Mixture Models (XDGMM) and Gaia DR2 data to characterize 420 previously reported clusters and discover 11 new clusters. They began by compiling a list of clusters from \cite{2007A&A...463..789A},  \cite{2002A&A...389..871D}, and \cite{2013A&A...558A..53K}. The candidates for each cluster came from a Gaia DR2 search with a target field of view centered on the cluster’s median position from these three studies and an opening angle equal to 1.5 times the largest cluster angular diameter from the same studies. Any targets with proper motion more than 10 sigma away from the cluster median were removed and only the 10,000 stars with the highest probability of being cluster members based off of their parallax were kept. Proper motions and parallax were then scaled to create an appropriate shape in parameter space before the XDGMM fit is performed. Each cluster was fit nine times with anywhere between 2 and 10 Gaussian components and the best fit was chosen according to the Bayesian Information Criterion. The differential entropy, a measure of the compactness of each Gaussian component, was then calculated and the component with the lowest differential entropy was designated as the cluster component. Individual membership probabilities were then calculated using bootstrap re-sampling on the selected Gaussian component. This procedure identified 601 $\alpha$ Per candidates. We note that the membership list from \cite{2021ApJ...923..129J} does not contain any unique candidates (i.e. all of the stars contained in Jaehnig’s membership list are also contained in the seven other studies included in our sample).

\subsection{Kerr et al. 2021} \label{subsec: kerr}

\cite{2021ApJ...917...23K} used Gaia DR2 and HDBSCAN to create the deepest and most comprehensive study of young stellar associations in the local neighborhood to date. The authors created their sample of Gaia DR2 objects by selecting all objects in Gaia DR2 with a parallax $>$ 3 mas and imposing quality cuts on unit weight error, BP/RP flux excess factor, visibility periods, and flux error to define a clean sample of candidates with reliable photometric measurements. In order to confirm that a star is young, a model population of 10 million stars was created that allows factors that could affect the photometric youth of a star, such as metallicity, multiplicity, and reddening, to be taken into account. Posterior distributions in mass and age were generated in the sample population corresponding to the locations of each star in the Gaia DR2 population, and the probability that each star is a young star was then estimated by integrating the age posterior over all ages less than 50 Myr. Stars with a probability of being a young star greater than 0.1 then underwent an additional parallax cut to create a final sample of 28,340 stars. The galactocentric XYZ positions and $v_l$ and $v_b$ tangential velocities for each star were then run through HDBSCAN. A total of 27 young associations were found in this population. HDBSCAN was then re-run on 10 of these populations that show visible substructure in order to characterize the hierarchical structures in each region. This process gave 1853 candidates in the $\alpha$ Per region.

\subsection{Meingast et al. 2021} \label{subsec: meingast}

\cite{2021A&A...645A..84M} used Gaia DR2 data to determine the morphology and dynamical structure of 10 nearby ($\leq$ 500 pc) and young (30–300 Myr) open star clusters with Gaia DR2 data, one of which is $\alpha$ Per. To start, the authors computed the mean cluster position and velocity of alpha per by using only stars from \cite{2018A&A...618A..93C} that were designated as $\alpha$ Per members and had membership probabilities greater than 0.8. All stars that fell within a large box (300 pc in X, 500 pc in Y, and 100 pc in Z) centered on the mean cluster position were then designated as candidates. Any candidate with a velocity greater than 1.5 km/s away from the bulk cluster velocity was filtered out and DBSCAN was used to extract overdensities in XYZ space from the remaining sample. The authors note that measurement errors in geometric distances along the line of sight resulted in elongated cluster structures. This issue was addressed by using extreme deconvolution to obtain deconcolved distances and, therefore, a more accurate shape for $\alpha$ Per. \cite{2021A&A...645A..84M} reported 1223 stars in $\alpha$ Per and estimated there to be 91 field star contaminants in their sample.

\subsection{Moranta et al. 2022} \label{subsec: moranta}

\cite{2022arXiv220604567M} used HDBSCAN with full 6-dimensional XYZ galactic positions and UVW space velocities of stars within 200 pc from Gaia EDR3 to identify 50 previously known associations, 32 new stellar streams, 9 extensions of groups recovered by \cite{2019AJ....158..122K}, and 8 new coronae. The authors defined their sample by selecting only stars in Gaia EDR3 with parallax above 5 mas and RUWE $<$ 3. The XYZ and UVW positions, velocities, and their respective uncertainties were then calculated with a Monte Carlo simulation and stars with uncertainties greater than 3 pc or 3 km\,s$^{-1}$ were rejected, creating a sample of 303,540 stars. The UVW velocities of each star in the sample were given the same dimensions as the XYZ positions through a transformation (multiplying each UVW velocity by 12 pc/km\,s$^{-1}$) before the stars were run through HDBSCAN, where the authors placed more importance on the kinematic distribution of the cluster than the spatial distribution. This procedure found 165 $\alpha$ Per candidates.

\section{Back-integrations at Young Ages} \label{sec: back int}

\begin{figure}[t]
    \centering
    \includegraphics[width=0.48\textwidth]{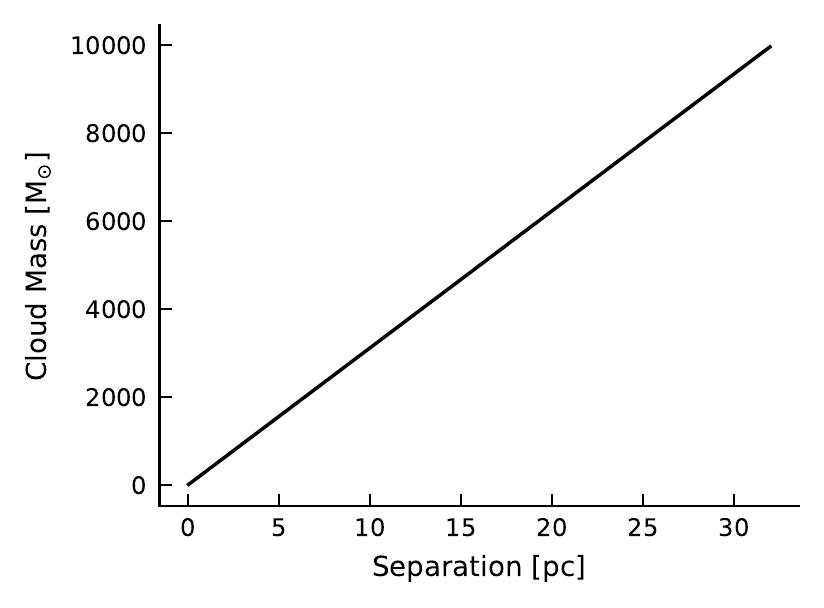}
    \caption{The mass needed to create a gravitational potential that overcomes the Milky Way's potential at $\alpha$ Per's position as a function of distance from the mass.}
    \label{fig:back int potentials}
\end{figure}

As can be seen in Figure \ref{fig:swoop plot}, $\alpha$ Per's core and the upper extension reached their minimum separation of 30 pc around 50 Myr ago. Beyond 50 Myr ago, the separation between the two regions increases. Why? When embedded clusters are born, the vast majority of mass in the cluster is contained in the gas itself, with only a few percent of the cloud's mass eventually being converted into stars. As the cluster ages, the gas that is not used in the star formation process is expelled. The model that we used for our back integrations only takes into account the Milky Way’s potential and does not take into account the gravitational potential of the gas from the early stages of the cluster’s life. The extra mass from the gas contributes to the cluster's gravitational potential and should overwhelm the gravitational potential of the Milky Way at close enough distances from the core, rendering the back integration inaccurate at the young ages when the core and upper extension were closest together.

To quantify this effect, we compared the potential of a point source with molecular cloud-like masses to the potential of the Milky Way at $\alpha$ Per's distance from the center of the Milky Way ($\sim8272$ pc). Both potentials were modeled in \texttt{galpy} with \texttt{MWPotential2014} for the Milky Way’s potential and \texttt{KeplerPotential} for the point source’s potential. \texttt{KeplerPotential} is the \texttt{galpy} implementation of a standard point source potential. For each mass in a range of masses from $\rm 10M_{\odot}$ to $\rm 10000M_{\odot}$, we calculated the point source’s potential and iterated outwards in distance until the point source’s potential was smaller than the Milky Way’s potential. The results of this procedure are displayed in Figure \ref{fig:back int potentials} and show that for the core and upper extension’s minimum separation of 30 pc, a $\rm \sim 9000 M_{\odot}$ cloud would be needed to overcome the Milky Way’s potential. This is roughly double the mass of the Taurus GMC \citep{1981MNRAS.194..809L} and, given that Taurus is known to have on the order of hundreds of members \citep[see e.g.][]{2019AJ....158..122K, 2021ApJ...917...23K} instead of thousands of members like $\alpha$ Per, this appears to be a very reasonable mass requirement.

\section{Selection Function} \label{sec: selection figure}

\begin{figure*}[t!]
    \centering
    \includegraphics[width=\textwidth]{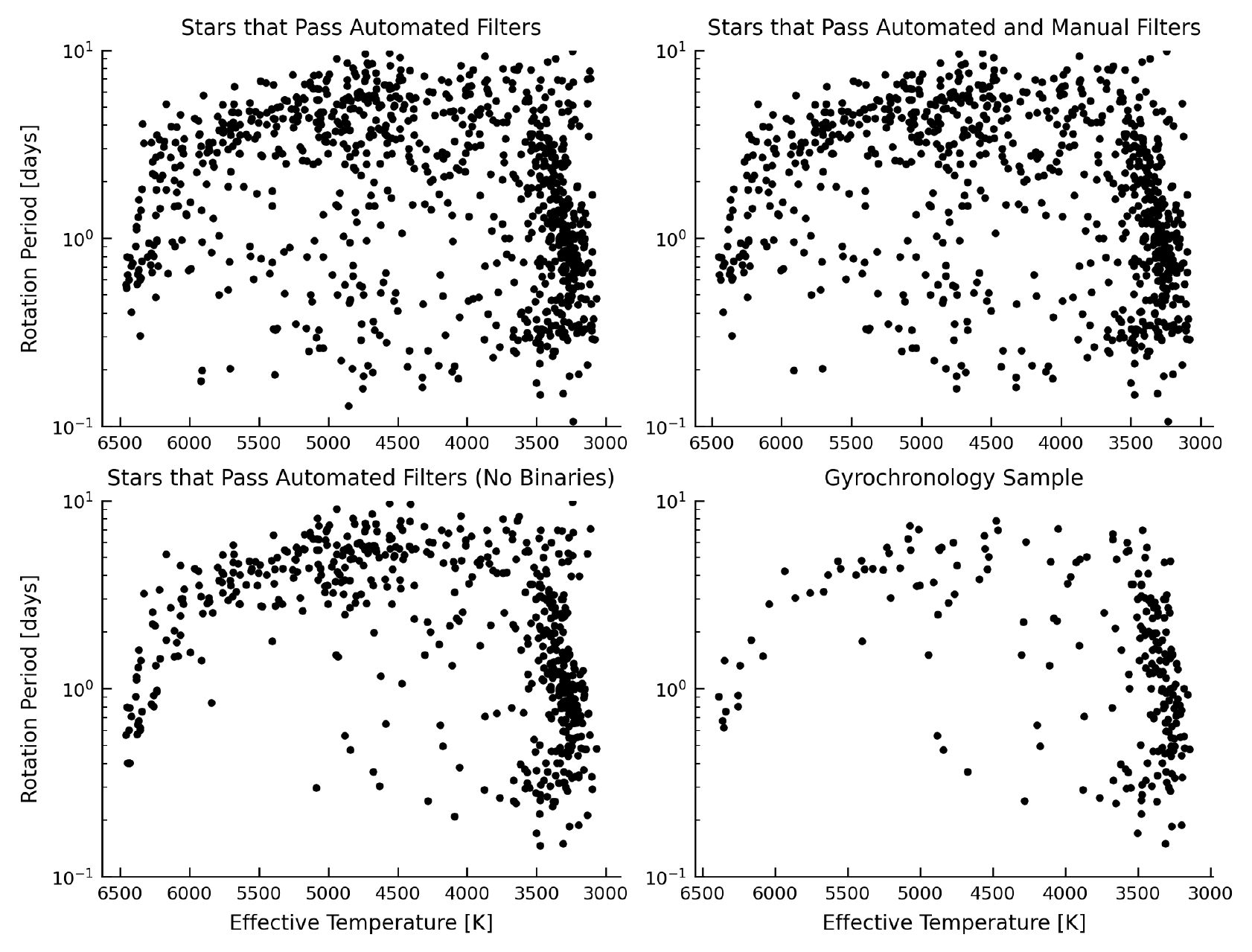}
    \caption{The rotation-effective temperature sequence for $\alpha$ Per with different choices of selection function. Upper left: $\alpha$ Per members selected using the brightness, internal consistency, periodogram strength, SNR, and external criteria described in Section \ref{sec: methods} (\texttt{flag\_quality\_period}). Upper right: $\alpha$ Per members selected using the same criteria as the upper left panel but that pass our manual check (\texttt{manual\_check} = g). Lower left: Same as upper left panel except with binaries removed according to the criteria described in Section \ref{sec:gyro} (\texttt{flag\_benchmark\_period}). Lower right: $\alpha$ Per members that comprise our sample of stars used to calibrate gyrochronology as described in Section \ref{subsec:gyro sample} (\texttt{in\_gyro\_sample}). This sample has the same binarity requirements as in the lower left panel, but only consists of stars that pass our manual vetting, have no sub-peaks in a periodogram within 70$\%$ the height of the main peak, and that lie within $\sim$\,30 pc and $\sim$3 km\,s$^{-1}$ of $\alpha$ Per's core.}
    \label{fig:selection plot}
\end{figure*}

We provide four different ways to select members of our $\alpha$ Per sample in Table 3:

\begin{enumerate}
    \item \texttt{flag\_quality\_period}: selects the 938 stars that pass our automated and manual quality checks on rotation period.
    \item \texttt{manual\_check} = g: selects the 863 stars that pass our manual vetting.
    \item \texttt{flag\_benchmark\_period}: selects the 593 stars that pass our binary checks defined in Section \ref{subsec:gyro sample}.
    \item \texttt{in\_gyro\_sample}: selects the 238 stars that are used to calibrate gyrochronology.
\end{enumerate}

Figure \ref{fig:selection plot} shows the effect of each of these selection functions in rotation-effective temperature space.

\section{Supplementary Figures} \label{sec: extra figures}

\begin{figure*}[t]
    \centering
    \includegraphics[width=\textwidth]{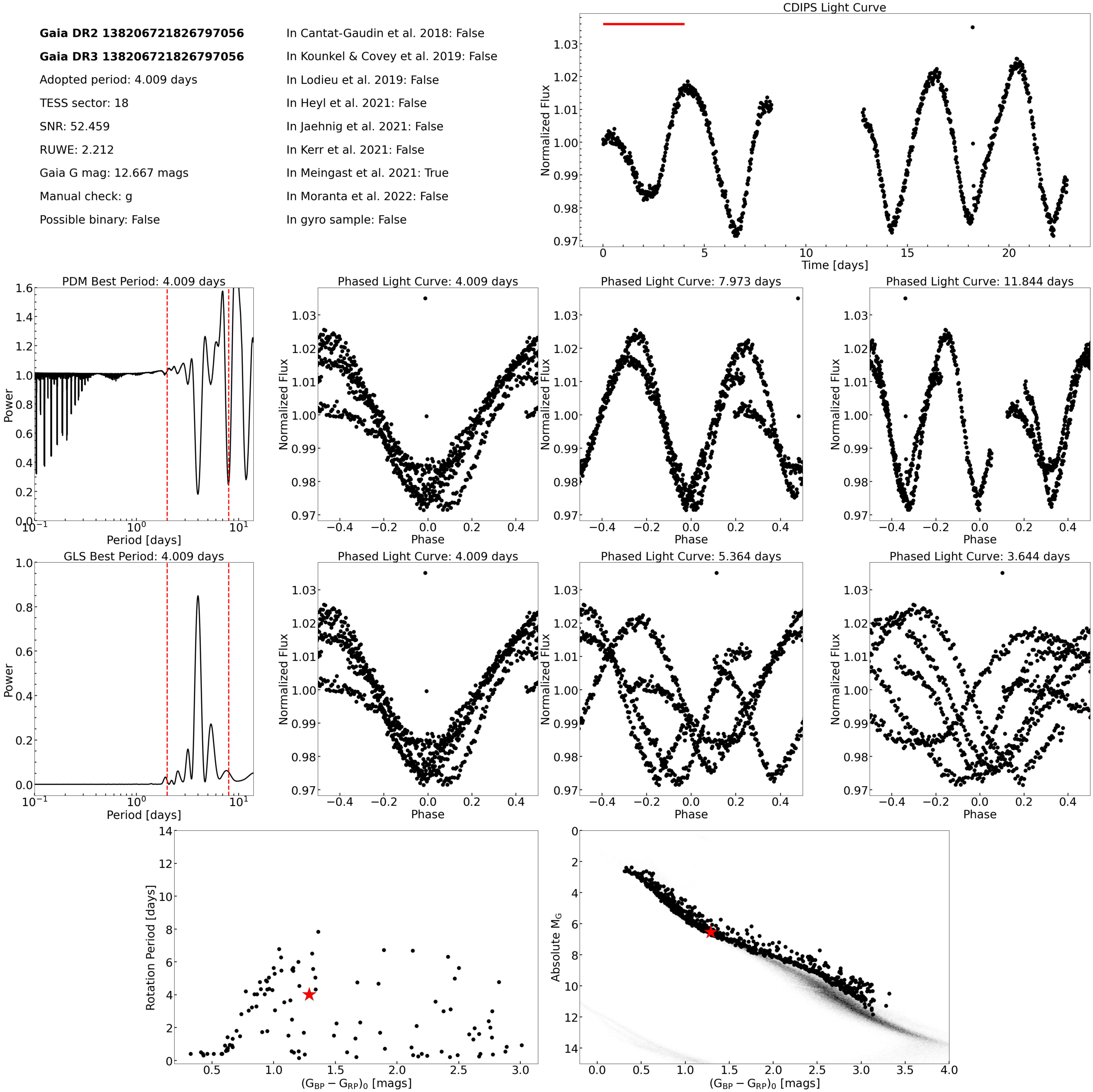}
    \caption{A diagnostic plot for Gaia DR2 138206721826797056. See Appendix \ref{sec: extra figures} for details. The complete figure set (986 diagnostic plots) is available in the online journal.}
    \label{fig:diagnostic plot}
\end{figure*}

We additionally include diagnostic plots for all 855 stars with a rotation period measurement that passed our automated quality checks and 131 additional stars that are included in our set of stars used to calibrate gyrochronology. An example is given in Figure~\ref{fig:diagnostic plot}. Basic information about each star is plotted in the upper-left panel. The CDIPS light curve is plotted in the upper-right panel with a horizontal red bar to indicate the rotation period. The GLS and PDM periodograms and light curves phase folded on the period from the three most prominent peaks in each periodogram are shown in the next rows. Dashed vertical red lines in the periodogram mark the location of double and half of the period measured by each method. The lower-left panel is the star's location on $\alpha$ Per's rotation--color diagram and the bottom-right panel is the star's location on a CAMD, with the grey points representing field stars and black points representing all 986 stars with a rotation period measurement that passed our automated quality checks or that was included in our gyrochronology sample.

\clearpage

\bibliography{sample631}{}
\bibliographystyle{aasjournal}



\end{document}